\documentclass[seceq,preprint]{ptptex}

\usepackage{graphicx}
\usepackage{color}
\preprintnumber[3cm]{YITP-06-68\\
 KEK-CP-188}
\title{%        %You can use \\ for explicit line-break 
Precision Study of $B^* B\pi$ Coupling for the Static Heavy-light Meson%
}

\author{%       %Use \scshape  for the family name
\ Shunsuke \textsc{Negishi},$^1$ %
Hideo \textsc{Matsufuru}$^2$ %
\ and  \ Tetsuya \textsc{Onogi}$^1$%
}

\inst{%     %Affiliation, neglected when [addenda] or [errata]
${}^1$Yukawa Institute for Theoretical Physics,
Kyoto University, \\
 Kyoto 606-8502, Japan\\
${}^2$ High Energy Accelerator Research Organization (KEK),\\
 Tsukuba 305-0801, Japan\\
}

\abst{%         %this abstract is neglected when [addenda] or [errata]
We compute the $B^*B\pi$ coupling $\hat{g}_{\infty}$ for the static
heavy-light meson using all-to-all propagators. It is shown that
low-mode averaging with 100 low-lying eigenmodes indeed significantly improves the
signal for the 2-point and 3-point functions for heavy-light meson.
 Our study suggests that the all-to-all propagator is
 a very efficient method for the high precision computation of the
$B^*B\pi$ coupling, especially in unquenched QCD, where the number of
configurations is limited.}

\begin{document}

\maketitle

%%%%%%%%%%%%%%%%%%%%%%%%%%%%%%%%%%%%%%%%%%%%%%%%%%%%%
%
\section{Introduction}
%
%----------------------------------------------------
Significant progress in B factory experiments has provided us with
crucial information to determine the Cabibbo-Kobayashi-Maskawa (CKM)
matrix elements and to test the standard model and the physics
beyond. Among all the components, the element $|V_{ub}|$ has attracted a great deal of
attention with regard to the problem of testing the consistency of the unitarity triangle when
combined with the angle $\sin(2\phi_1)[=\sin(2\beta)]$. The consistency 
of the constraints derived from $|V_{ub}|$ and $\sin(2\phi_1)$ is especially
interesting because $|V_{ub}|$ is determined by a tree 
level decay process, while the decay that determines $\sin(2\phi_1)$  has
contributions from penguin amplitudes, which is sensitive to new
physics. Although  $\sin(2\phi_1)$ is already known with a few percent
accuracy, $|V_{ub}|$ is known only within 10\% -- 20\% from the inclusive 
and the exclusive semileptonic B decays. Moreover, the constraint on the
unitarity triangle given  by $|V_{ub}|$ from the inclusive decay is only
marginally consistent with that from $\sin(2\phi_1)$. Therefore, it is
 important to reduce the error in the determination of $|V_{ub}|$ from
the exclusive decay, which can only be realized by
improving the form factor calculation in lattice QCD.   

In general, extracting the form factors in lattice QCD is numerically
more difficult than extracting decay constants or bag parameters, since it involves
3-point meson correlators with nonzero recoil momenta, whose treatment generally introduces
 non-negligible statistical and discretization
errors. However, in order to determine the CKM matrix elements, such as 
$|V_{ub}|$ or $|V_{cb}|$, one does not need to
know the form factor over a finite momenta, and in fact even the form factors
at a single value of  $q^2$ are sufficient. Thus, in principle, the form
factors at zero momentum recoil can determine the CKM matrix
elements, provided that the
experimental data are statistically sufficient. This is indeed the case for the determination of $|V_{cb}|$
from the $B\rightarrow D l\nu$ process. The situation for $B\rightarrow\pi l\nu$
also seems promising, because there has been a rapid improvement in the quality of the experimental data that can be obtained.
 In fact, recently the BaBar collaboration observed the $q^2$
dependence of the $B\rightarrow\pi l\nu$ decay quite
precisely~\cite{Babar_Hawaii}.    

It is known that symmetries often can be used to greatly simplify 
problems. In the case of $B$ mesons, the approximate chiral and heavy
quark symmetries restrict the hadronic amplitudes. In particular, the
effective Lagrangian, which has both chiral and heavy
quark symmetries, together with small symmetry breaking effects, allows us to
obtain explicit relations among decay constants and form factors through
 systematic expansions in chiral perturbation theory and in terms of 
$1/M$, where $M$ is the heavy quark mass.  An important example is provided by
the form factors of $B\rightarrow\pi l\nu$ semileptonic decay. By the
chiral symmetry, these form factors can be expressed in terms of the heavy-light
decay constants  and the coupling of the vector and pseudoscalar
heavy-light mesons to the pion, $g_{B^* B\pi}$. In general the pionic
coupling of  heavy-light meson is defined as
\begin{eqnarray}
g_{H^* H\pi} = \frac{2 \sqrt{m_H m_{H^*}}}{f_{\pi}} \hat{g}_Q, 
\end{eqnarray}
where $\hat{g}_Q$ is the dimensionless coupling that appears
in the heavy meson effective theory.
Therefore, we can reduce the problem of
computing the form factor to simpler problems of computing the decay
constant and the coupling $\hat{g}_b$.

The main sources of uncertainties in the types of computations considered here are
(1) the quenching error,  (2) the chiral   extrapolation error for
light quarks, and (3) the discretization and/or perturbative errors
from  heavy quarks. The first two of these are common to almost all
quantities, and unquenched QCD simulations with quarks that are sufficiently light
 to solve these problems are being actively studied with various
lattice actions. To reduce the third type of error, however, formulations of
the lattice heavy quark which allow the nonperturbative
renormalization and the continuum limit are necessary. Promising
approaches have been proposed by the Alpha
collaboration~\cite{Heitger:2003nj}, the Rome II
group,\cite{deDivitiis:2003wy}$^,$~\cite{deDivitiis:2003iy} and their
joint collaboration~\cite{Guazzini:2006bn}. 
In this approach, one uses the nonperturbative heavy quark effective theory (HQET)
through order $1/M$ or relativistic QCD with a finite size scaling
technique, or combinations of both.  
It has been found that in the first and third approaches, the results in
the static limit play a crucial role in precise computation of
the $B$ meson.   

For the above reasons, there have been several calculations of the
$B^* B\pi$ coupling using HQET in both quenched and unquenched
lattice QCD. In order to use $B^*B\pi$ coupling for the precise
determination of 
$|V_{ub}|$, the sea quark effect from the unquenched calculation and
the $1/M$ correction are essential. The major drawback of HQET is that the
static 
propagators are much noisier. This limits the accuracies and also makes
it difficult to study $1/M$ corrections. In the quenched case, one can
make use of a large number of gauge configurations to reduce the statistical
error, while in the unquenched case, clever techniques to reduce the  statistical
error for a limited number of configurations are necessary. 

The Alpha collaboration~\cite{DellaMorte:2003mn} proposed
 using a smeared link for the HQET action, 
which can reduce the noise/signal  ratio significantly. In
particular, they showed that the HYP smearing~\cite{Hasenfratz:2001hp}
is the most efficient.

Low-mode averaging is also known to improve the statistical accuracy.
The TrinLat group~\cite{Foley:2005ac} proposed a new method employing an all-to-all propagator
 by
combining the low mode averaging and the noisy estimation. They showed 
that their all-to-all propagator significantly improves the
heavy-light meson correlators, at least on coarse lattices.

In this paper, we aim at a high precision computation of the coupling 
 $\hat{g}_{\infty}$. 
With the goal of realizing high precision computations of the $B^* B\pi$ coupling in dynamical
simulations, we carry out a systematic feasibility study on quenched gauge
configurations, combining the two improved techniques of the HYP
smeared link and the all-to-all propagators with low-mode averaging.    
We find that 100 low eigenmodes dominate the 2-point and 3-point
static-light correlators after a few time slices. This 
implies that the low-mode averaging technique
indeed does help to reduce the error significantly. Although the same error reduction
is achieved by increasing the number of configurations in the quenched
case, it has a big advantage in the unquenched case, where the 
number of configurations by the computational resources.

This paper is organized as follows. In \S 2, we summarize the form
factors of the $B\rightarrow\pi l\nu$ decay and the implications of recent
experiments with regard to the determination of $|V_{ub}|$. In \S 3, we explain 
the soft pion relations of the form factors using the chiral
perturbation theory  for heavy-light mesons. In \S 4, we review
the two techniques to improve the precision, i.e. those employing the lattice
HQET action 
with the HYP smeared link and all-to-all propagators. In \S
5, we explain the methods to compute the coupling
$\hat{g}_b$. Details of the simulation are given in \S 6. We present our
numerical results in \S 7. Conclusions are given in \S 8. 

%%%%%%%%%%%%%%%%%%%%%%%%%%%%%%%%%%%%%%%%%%%%%%%%%%%%%
%
\section{Form factor}
%
%----------------------------------------------------

The form factors for $B\rightarrow \pi l\nu$ decay are defined as
\begin{eqnarray}
  \label{eq:f+f0}
  \langle\pi(k_{\pi})|\bar{u}\gamma^{\mu}b|B(p_B)\rangle
&  = & f^+(q^2) 
  \left[ 
    (p_B+k_{\pi})^{\mu} 
    - \frac{m_B^2-m_{\pi}^2}{q^2} q^{\mu}
  \right] 
\nonumber\\
& &   + f^0(q^2) \frac{m_B^2-m_{\pi}^2}{q^2} q^{\mu}, 
\end{eqnarray}
where the momenta $p_B$ and $k_{\pi}$ are, respectively those of the $B$
and $\pi$ mesons and $q = p_B-k_{\pi}$ is the momentum transfer. The value of $q^2$
ranges from 0 to $q_{\rm max}^2=(m_B-m_{\pi})^2$. Ignoring the lepton
mass, the branching fraction of the semileptonic decay $B\rightarrow
\pi l\nu$ is given in terms of $f^+(q^2)$ only by  
\begin{eqnarray}
\Gamma(B\rightarrow\pi l\nu)
&=& \frac{G_F^2 |V_{ub}|^2}{24\pi^3}  
\left [\frac{(q^2 -(m_B+m_{\pi})^2)(q^2 -(m_B-m_{\pi})^2)}
{(2m_B)^2}\right]^{3/2}
\int dq^2 |f^+(q^2)|^2. \nonumber\\
\label{eq:DecayRate}
\end{eqnarray}
As seen from this equation, the hadron matrix elements
$\langle\pi(k_{\pi})|\bar{u}\gamma^{\mu}b|B(p_B)\rangle$ are given
by a linear combination of the form factors $f^+(q^2)$ and
$f^0(q^2)$, and hence one must solve the linear equation for two
independent matrix elements in order to obtain the form factors. 
Because one of the matrix elements vanishes at zero recoil, the
numerical accuracy of $f^+(q^2)$ near the exact zero recoil is
extremely poor. For this reason, the extraction of the form factors 
is based on the matrix elements with small but nonzero recoil
momentum. This gives rise to somewhat large statistical and
discretization errors.  
In fact, previous lattice calculations in quenched QCD
~\cite{Bowler:1999xn}$^-$~\cite{Aoki:2001rd}
% ~\cite{Abada:2000ty} ~\cite{El-Khadra:2001rv}
 and in unquenched QCD~\cite{Gulez:2006dt}$^-$~\cite{Shigemitsu:2004ft}
%~\cite{Okamoto:2004xg}
 have yielded determinations of
form factors with typically 20\% accuracies, and it seems difficult to
reduce the errors with the present technique.  Fortunately, at
zero recoil, one can use the soft pion relation to relate the form
factor to simpler quantities, i.e. the decay constants and the
$B^*B\pi$ coupling,
\begin{eqnarray}
  \label{eq:pole_dominance_f+}
  \lim_{q^2\rightarrow m_{B^*}^2} f^+(q^2) =
  \frac{f_{B^*}}{f_{\pi}}
  \frac{\hat{g}}{1-q^2/m_{B^*}^2}.
\end{eqnarray}
Since the hyperfine splitting of $B^*$ and $B$ is 45 MeV,
the direct decay of $B^* \rightarrow B \pi$ cannot be measured 
experimentally, but it can be computed using lattice QCD 
quite accurately. A quenched lattice computation of
$\hat{g}_{\infty}$ with HQET was first carry out by the UKQCD collaboration,
~\cite{McNeile:2004cb} 
and subsequently Abada et al~\cite{Abada:2003un}.
 made a more precise calculation. 
Becirevic et al. also  computed
$\hat{g}_{\infty}$ in $n_f=2$ unquenched QCD~\cite{Becirevic:2005zu}. The currents value of $\hat{g}_\infty$ are
\begin{eqnarray}
\hat{g}_{\infty} &=& 0.51\pm 0.03_{\rm stat}\pm 0.11_{\rm sys}\mbox{  for $n_{f}=0$},\\
\hat{g}_{\infty} &=& 0.51\pm 0.10_{\rm stat}\hspace{1.7cm} \mbox{  for $n_{f}=2$}.
\end{eqnarray}
It should be noted that the quenched result is quite accurate and once
one succeeds in obtaining the $1/M$ correction for the $B$ meson either by
interpolating the results in the static limit and the charm quark
region or by computing the $1/M$ correction in HQET, the form
factors at zero recoil can be predicted precisely. However, in unquenched QCD, 
the statistical error is one of the major sources of uncertainty.
Given this situation, some clever techniques are needed to improve the
accuracy with limited statistical samples. 

The experimental situation is quite interesting. 
The Belle and BaBar collaborations are measuring $B\rightarrow\pi l\nu$, and
they have even 
measured the $q^2$ dependence. In the most recent results,  BaBar
measured the partial branching fractions  of $B\rightarrow\pi l\nu$ 
decay with twelve energy bins for $q^2$ of size 2 GeV$^2$, where 
$q$ is the momentum of the lepton pair:
\begin{eqnarray}
\Delta \Gamma(q_i^2<q^2<q_{i+1}^2).  \mbox{ \ \  }(i=1,\cdots,12)
\end{eqnarray}
We should be able to extrapolate the experimental data to $q^2_{\rm max}$ in the
near future. Although the present experimental statistical error is
still large, the super B factory is expected to produce 50-100
times more data, which would allow a precise determination 
of the differential decay rate in the soft pion limit, similarly to
that for $B\rightarrow D l\nu$ decay.
 
%%%%%%%%%%%%%%%%%%%%%%%%%%%%%%%%%%%%%%%%%%%%%%%%%%%%%
%
\section{Chiral perturbation theory for the heavy-light meson}
%
%----------------------------------------------------
The low energy dynamics of the heavy-light meson and the light
pseudoscalar meson system can be
described systematically using the effective meson theory
incorporating 
the breaking terms of the chiral and  the heavy quark symmetries.
Defining the chiral field as $\xi=\exp(i{\cal M}/f)$ with the light
pseudoscalar meson field $\cal M$ and the heavy meson field $H$, which is defined in terms of the
heavy-light pseudoscalar and vector fields $B$ and $B^*$ as
\begin{eqnarray}
H   &\equiv& \frac{1}{2} (1+\gamma^{\mu} v_{\mu}) 
[i B \gamma_5 + B_i^* \gamma_i],
\end{eqnarray}
where the heavy meson field transforms linearly under the chiral and
heavy quark symmetry transformations. 
The light-meson part is the ordinary chiral Lagrangian and the
heavy-light meson part of the effective Lagrangian is  
\begin{eqnarray}
L &=& - \mbox{Tr}[\bar{H}iv \cdot DH]
+ g \mbox{Tr}[\bar{H}H {\cal A}_{\mu}\gamma^{\mu}\gamma_5]
+ O(1/M),
\end{eqnarray}
where ${\cal A}_{\mu}= \frac{i}{2} 
(\xi^{\dagger} \partial_{\mu} \xi
 -\xi \partial_{\mu} \xi^{\dagger})$, and $v$ is the velocity of the heavy meson. 
The pionic coupling of the heavy-light meson appears in the leading
term of the Lagrangian. After including $O(1/M)$ correction, the
effective coupling is   
$g_{B^*B\pi}= g + \displaystyle{\frac{g_1-g_2}{M}}$, 
where $g_1$ and $g_2$ are the coefficients in the next-to-leading term of the
Lagrangian.
The heavy-light V-A current is given by 
\begin{eqnarray}
(V-A)^{hl}_{\mu}= \frac{i\alpha}{2} \mbox{Tr}[\gamma_{\mu}(1-\gamma_5)
H \xi^{\dagger} ] + O(1/M),  
\end{eqnarray}
where $\alpha = f_B \sqrt{m_B}$.
Using this Lagrangian, it is straightforward to derive the soft pion 
relation systematically, including the corrections in the chiral
perturbation theory, as well as $1/M$ corrections. 
At leading order in the chiral perturbation theory, and through 
order $1/M$ in the heavy quark expansion~\cite{Boyd:1994pa}, we have
\begin{eqnarray}
f^+(q^2)
&=& -\frac{f_{B^*}}{2f_{\pi}} 
\left[ g_{B^*B\pi} (\frac{1}{v\cdot k_\pi -\Delta} - \frac{1}{m_B}) 
+\frac{f_B}{f_{B^*}}\right],
\end{eqnarray}
where $\Delta\equiv m_{B^*}-m_B$.

The $B^*B\pi$ coupling can be obtained from the form factor
corresponding to the matrix element 
\begin{eqnarray}
\langle B^*(p_{B^*},\lambda) | A^{\mu} | B(p_B) \rangle
&=& 2 m_{B^*} A_0(q^2) \frac{\epsilon^{\lambda}\cdot q}{q^2}q^{\mu}
+  (m_{B^*}+m_B) A_1(q^2)  
   \left[ \epsilon^{\lambda~\mu} - \frac{\epsilon^{\lambda}\cdot
q}{q^2}q^{\mu} \right] 
\nonumber\\ 
& & + 
    A_2(q^2) \frac{\epsilon^{\lambda}\cdot q}{m_{B^*}+m_{B}}
   \left[ p_{B^*}+p_{B} -\frac{m_{B^*}^2-m_B^2}{q^2} q^{\mu} \right],
\end{eqnarray}
where $p_B$ and $p_{B^*}$ are the momenta of $B$ and $B^*$ and we have
$q=p_B-p_{B^*}$. The $B^* B\pi$ coupling can be extracted 
from the residue of the pion pole with longitudinal polarization
as 
\begin{eqnarray} 
g_{B^* B\pi} 
&=& \frac{2 m_{B^*} A_0(0)}{f_{\pi}} 
= \frac{1}{f_{\pi}} 
    [ (m_{B^*}+ m_B) A_1(0) + (m_{B^*} - m_B) A_2(0) ],
\end{eqnarray} 
where we have used the kinematical constraint that 
the matrix element does not diverge at $q^2=0$. 
Thus the pionic coupling $\hat{g}_Q$ is given by 
\begin{eqnarray} 
\hat{g}_Q
&=&  \frac{m_{B^*} + m_B}{2\sqrt{m_B m_{B^*}}} A_1(0) 
    +  \frac{m_{B^*} - m_B}{2\sqrt{m_B m_{B^*}}} A_2(0) .
\label{eq:g_vs_A1_A2}
\end{eqnarray} 
In the static limit, Eq.(\ref{eq:g_vs_A1_A2}) simplifies to 
\begin{eqnarray} 
\hat{g}_Q &=& \hat{g}_{\infty}= A_1(0) .
\end{eqnarray} 
Now, the form factor $A_1(0)$ in the static limit can be obtained from
the matrix element at zero recoil as  
\begin{eqnarray}
\frac{\langle B^* | A_i(0) | B \rangle}{2m_B}
&=& A_1(0)  \epsilon_i^{\lambda},
\end{eqnarray}
which can be evaluated using lattice QCD calculations.
%
%%%%%%%%%%%%%%%%%%%%%%%%%%%%%%%%%%%%%%%%%%%%%%%%%%%%%
%
\section{Lattice HQET and all-to-all propagators}
%
%----------------------------------------------------
The lattice HQET action in the static limit is defined as
\begin{eqnarray}
S &=& \sum_x\bar{h}(x) \frac{1+\gamma_0}{2}
  \left[ h(x) - U_4^{\dagger}(x-\hat{4})h(x-\hat{4})  \right],
\end{eqnarray}
where $h(x)$ is the heavy quark field. The heavy quark propagator 
is obtained by solving the evolution equation derived from the action
with negligible numerical cost. Despite the advantages of being numerically economical
 and allowing better control over the systematic errors than other
lattice heavy quark formulations, the lattice calculations of heavy-light 
systems with HQET suffer from large statistical noise. The reason for this noise can be
understood as follows. First, consider the 2-point function of a heavy-light
meson. The signal $C_2(t)$ and the noise $\Delta C_2^2(t)$ behave as
\begin{eqnarray}
C_2(t) 
&=& \langle [\bar{Q}q](t) [\bar{q}Q](0) \rangle 
\sim \exp(-E_{Qq}t),
\\
\Delta C^2_2(t) 
&=& \langle |[\bar{Q}q](t) [\bar{q}Q](0)|^2 \rangle 
\sim \exp\left[-(E_{\bar{Q}Q}+E_{\bar{q}q})t\right] .
\end{eqnarray}
Thus, the noise-to-signal ratio is
\begin{eqnarray}
\frac{\rm noise}{\rm signal}
\sim \exp\left[(E_{\bar{Q}q}-\frac{E_{\bar{Q}Q}+E_{\bar{q}q}}{2})t\right] .
\end{eqnarray}
In the static limit, the corrections to the self-energy and binding energy 
of the heavy-heavy system exactly vanish, while in the heavy-light
system, the power divergent self-energy correction gives a significant 
contribution to the energy. Therefore, the noise-to-signal ratio is 
\begin{eqnarray}
\frac{\rm noise}{\rm signal}
\sim \exp\left[(\delta m_Q +E^{\rm
binding}_{\bar{Q}q}+\frac{E_{\bar{q}q}}{2})t\right], 
\end{eqnarray}
where the mass shift has the regularization dependent power divergent 
term $\delta m_Q \sim c/a$, with $c$ being the
regularization dependent constant. This implies that the noise
problem becomes increasingly serious as the continuum limit is approached.

In order to reduce the noise, the Alpha collaboration proposed a new HQET
action~\cite{DellaMorte:2003mn},
 in which the link variables are smeared to suppress the power
divergence. They studied the static heavy-light meson in the cases that the links are smeared
by APE smearing, HYP smearing~\cite{Hasenfratz:2001hp}
and a one-link integral and found that
the smeared link significantly reduces the noise. This result is indeed
consistent with the observation that the power divergent mass shifts at
one loop are reduced by smearing. For a $\beta=6.0$ lattice with 5000
configurations, the noise-to-signal for the time extent of 1-2 fm
remains at the 0.5 -- 1\% level. This means that using smeared links and 
data with high statistics allows precise computation of the static-light meson.
Indeed, Abada et al~\cite{Abada:2003un}.
used the HYP smeared link action and
obtained  $\hat{g}_{\infty}$ with 160 configuration to 3--5\%
accuracy, while for the unquenched case with 50 configurations, the
statistical error of $\hat{g}_{\infty}$ was on the order of 10\%.

Another useful technique to reduce the statistical error is to employ the
all-to-all propagators using low-mode  averaging. Low-mode averaging
was introduced by DeGrand et al~\cite{DeGrand:2002gm}.
for the overlap fermion and
has been used extensively in the $\epsilon$-regime ~\cite{Giusti:2004yp}, along with baryon
propagators~\cite{Giusti:2005sx}. 
The TrinLat~\cite{Foley:2005ac}
 collaboration developed a comprehensive method using
all-to-all propagators by combining the low-mode averaging and noisy estimator
with time, color, and spin dilutions. Their method is as follows. 
First, define the lattice Dirac Hamiltonian $Q$ as 
\begin{eqnarray}
Q = \gamma_5 D,
\end{eqnarray}
where $D$ is the lattice Dirac operator.
Then, the quark propagator $S_q(x,y)$ can written in terms of the inverse of 
the Hamiltonian $\bar{Q}=Q^{-1}$ as 
\begin{eqnarray}
S_q(x,y) = \bar{Q}(x,y) \gamma_5.
\end{eqnarray}
The hermitian operator $Q$ can be decomposed into low-mode and 
high-mode parts by making use of the spectral decomposition as
\begin{eqnarray}
Q  = Q_0+Q_1, &&\\
Q_0 =\sum_{i=1}^{N_{\rm ev}} \lambda_i v^{(i)} \otimes v^{(i) \dagger},\  
&& Q_1=\sum_{j=N_{\rm ev}+1}^N \lambda_j v^{(j)} \otimes v^{(j) \dagger},
\end{eqnarray}
where $\lambda_i$ is the eigenvalue associated with the eigenvector $v^{(i)}$, 
with the index labeling eigenmodes in the order of increasing absolute value of 
the eigenvalue. Correspondingly, the propagator $\bar{Q}$ is also
decomposed as  
\begin{eqnarray}
\bar{Q}   = \bar{Q}_0+\bar{Q}_1,&&\\
\bar{Q}_0 = \sum_{i=1}^{N_{\rm ev}} \frac{1}{\lambda_i}
 v^{(i)} \otimes v^{(i) \dagger},\ 
&& \bar{Q}_1 = \sum_{j=N_{\rm ev}+1}^N \frac{1}{\lambda_j}
 v^{(j)} \otimes v^{(j) \dagger}
         = \bar{Q}P_1, 
\end{eqnarray}
where
\begin{eqnarray}
P_1 = 1 - P_0 = 1 - \sum_{i=1}^{N_{\rm ev}}
 v^{(i)} \otimes v^{(i) \dagger}
\end{eqnarray}
is the projection operator that projects into the space consisting of only the modes above
the $N_{\rm ev}$-th lowest eigenmodes. Determining the low-lying 
eigenmodes, one can compute $\bar{Q}_0$ directly and evaluate the remaining 
high-mode part, $\bar{Q}_1$, with a noisy estimator.

Generating a random noise vector $\eta$ which satisfies the
property
\begin{eqnarray}
\langle \langle \eta(x)^a_\alpha \eta(y)^{*b}_\beta \rangle \rangle
\equiv 
\lim_{N_r\rightarrow \infty}
\frac{1}{N_r}\sum_{[r]}
\eta_{[r]}(x)^a_\alpha \eta_{[r]}(y)^{*b}_\beta
= \delta_{xy}\delta_{ab}\delta_{\alpha\beta},
\label{eq:noise}
\end{eqnarray}
and introducing $\psi_{[r]}(x)$ with 
\begin{eqnarray}
\psi_{[r]}(x) = \sum_y(\bar{Q}P_1)(x,y)\eta_{[r]}(y),
\end{eqnarray}
we can obtain the all-to-all propagator as
\begin{eqnarray}
\langle\langle 
\psi_{[r]}^a(x)_\alpha\otimes\eta_{[r]}^{\dagger}(y)^b_\beta 
\rangle\rangle
=(\bar{Q}P_1)(x,y)^{ab}_{\alpha\beta},
\end{eqnarray}
where we have used the property in Eq.~(\ref{eq:noise}).
The TrinLat collaboration~\cite{Foley:2005ac}
showed that the dilution is efficient for decreasing
the statistical error of the correlator evaluated with noisy
estimator. The dilution consists of the decomposition of the noise vector $\eta$
into each time, spin and color component as
\begin{eqnarray}
\eta_{[r]}(\vec{x},t)^a_\alpha = 
\sum_{j} \eta_{[r]}^{(j)}(\vec{x},t), 
\end{eqnarray}
where $j$ is the index for the dilution labeling the time, spin, and
color sources, i.e. $j=(t_0,\alpha_0,a_0)$, and $\eta_{[r]}^{(j)}(\vec{x},t)$ is defined by
\begin{eqnarray}
\eta^{(j)}_{[r]}(\vec{x},t)^a_\alpha
 =  \eta_{[r]}(\vec{x})^a_\alpha 
\delta_{t, t_0} \delta_{a, a_0} \delta_{\alpha, \alpha_0}.
\end{eqnarray}
After applying the dilution, the all-to-all propagator reads
\begin{eqnarray}
(\bar{Q}P_1)(x,y)
 = \lim_{N_r\rightarrow\infty}
\sum_r^{N_r}\sum_j \psi_{[r]}^{(j)}\otimes\eta^{\dagger}(x)_{[r]}^{(j)}.
\end{eqnarray}

The TrinLat group showed that their all-to-all propagator is indeed 
useful for obtaining a good signal for the heavy-light system on coarse
lattices. Although their proposal seems promising, because the power
divergence becomes increasingly severe as the lattice becomes finer, the magnitude of the quantitative
improvement is realized for finer lattices  remains to be
investigated.  

%%%%%%%%%%%%%%%%%%%%%%%%%%%%%%%%%%%%%%%%%%%%%%%%%%%%%
%
%hm \section{Calculational methods}
\section{Calculational methods}
%
%----------------------------------------------------
The matrix element $\langle B^* | A_{\mu} | B \rangle $
%hm  in the static limit 
at zero recoil
can be obtained from the ratio of the 3-point and 2-point correlation
functions $R(t)$ as 
%\begin{eqnarray}
\begin{equation}
\frac{ \langle B^*(0)| A_i | B(0) \rangle }{2 m_B} 
= \lim_{t, t_A \rightarrow \infty} R(t,t_A), \\
\end{equation}
%\mbox{where}
where
\begin{equation}
R(t,t_A) = 
\frac{\langle {\cal O}_{B^*}(t+t_A) A_i(t_A) {\cal O}_B(0) \rangle }
{\langle {\cal O}_{B^*}(t+t_A) {\cal O}_B(0) \rangle },
\end{equation}
%\end{eqnarray}
%hm with ${\cal O}_{B,B^*}$ are some smeared operator having quantum numbers 
with the operators ${\cal O}_{B}$ and ${\cal O}_{B^*}$ having the quantum
numbers of the $B$ and $B^*$ mesons, respectively.
We apply the smearing technique to enhance the ground state contributions
to the correlators as
\begin{eqnarray}
{\cal O}_B(t,\vec{x}) 
 &=& \sum_{\vec{r}} \phi(\vec{r})\bar{q}(t,\vec{x}+\vec{r})
   \gamma^5 h(t,\vec{x}),\\ 
{\cal O}_{B^*}^i(t,\vec{x})
 &=& \sum_{\vec{r}} \phi(\vec{r})\bar{q}(t,\vec{x}+\vec{r})
  \gamma^i h(t,\vec{x}),
\end{eqnarray}
where $\phi(\vec{x})$ is the smearing function.
%The 2-point and 3-point correlation functions
%are divided into higher and lower parts as below. 
To apply the all-to-all propagator technique introduced in the
last section, we decompose the 2-point and 3-point correlation
functions into the high-mode and low-mode parts as described below.

\begin{table}[tb]
\caption{The quantum numbers of the mesons and
the corresponding gamma matrices.}
\begin{center}
\begin{tabular}{ccccc}
\hline
   &  $\Gamma$ & $\gamma_5\Gamma$ &
      $\tilde{\Gamma}_5$ & $\tilde{\Gamma}_5\gamma_5$ \\
\hline
Ps  & $\gamma_5$ & 1 & $-1$ &  $-\gamma_5$ \\
V   & $\gamma_j$ & $\gamma_5\gamma_j$ &
              $-\gamma_5\gamma_j$ &  $-\gamma_j$ \\
\hline
\end{tabular}
\end{center}
\label{tab:gamma}
\end{table}

\begin{figure}[tb]
  \begin{center}
\vspace{-0.2cm}
      \includegraphics[width=6.0cm]{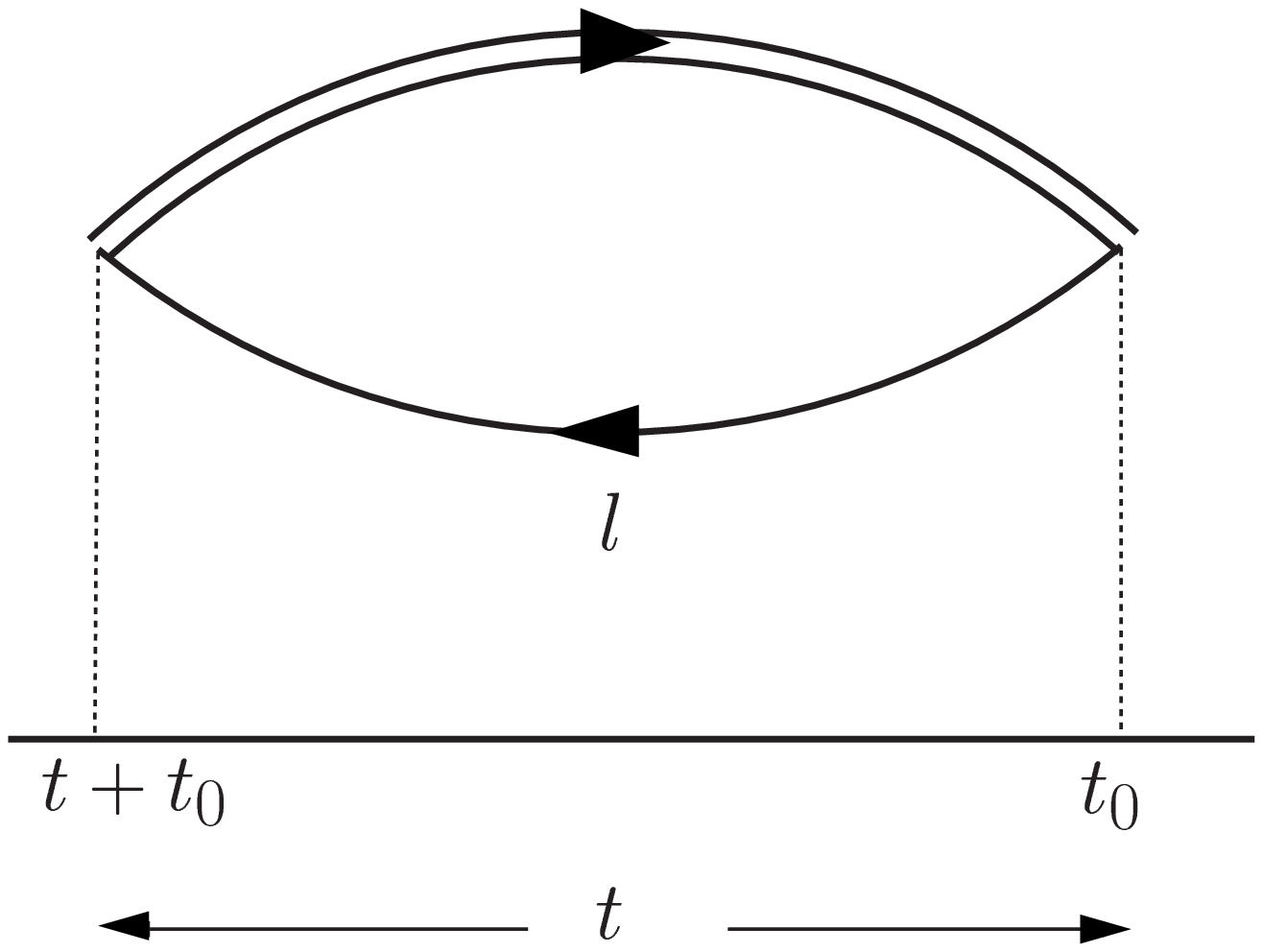}
\hspace{1cm}
      \includegraphics[width=6.0cm]{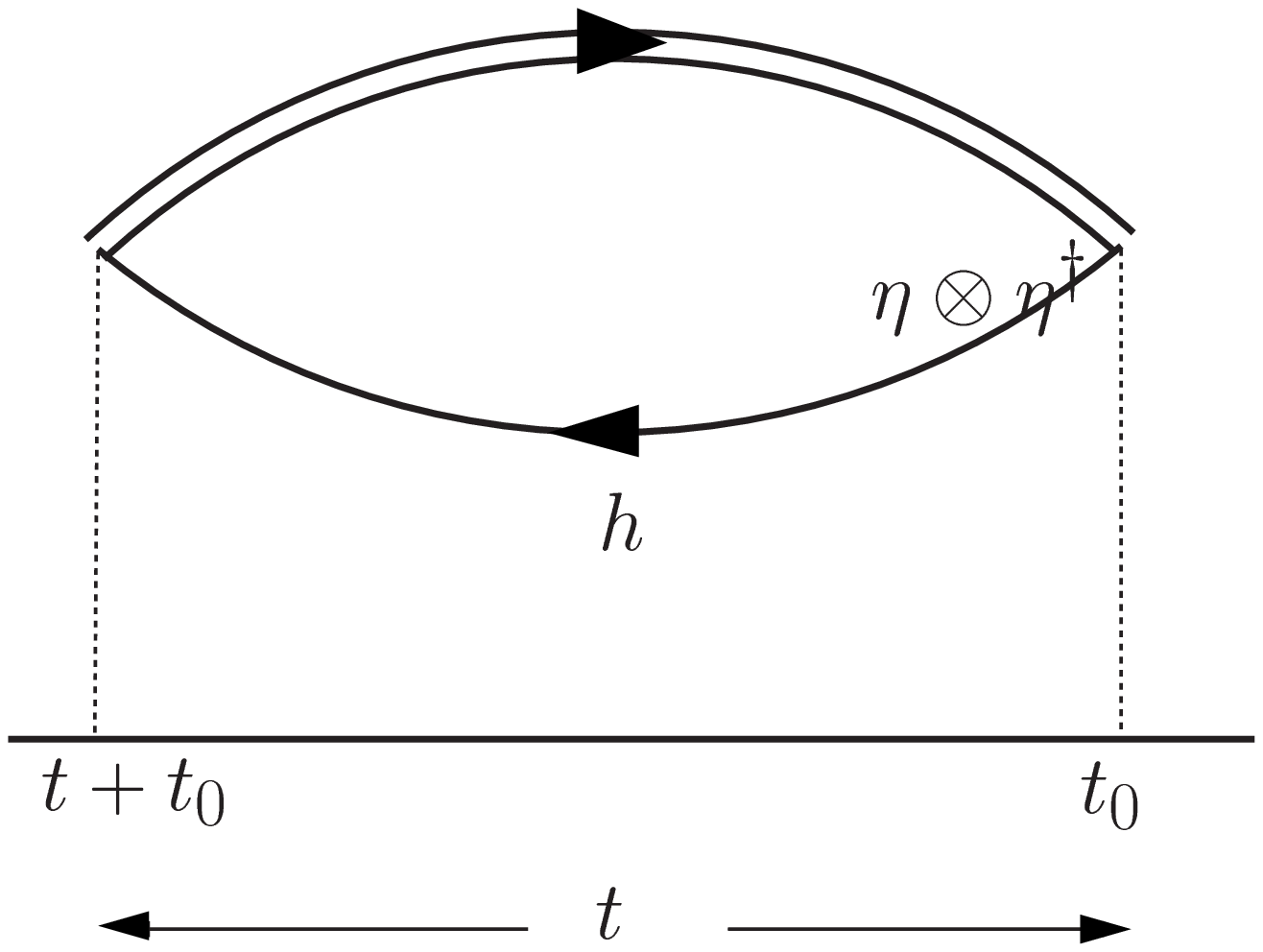}
%\vspace{-5.5cm}
  \end{center}
   \caption{The diagrams for the 2-point correlation function. The left
   and right panels show the low-mode and high-mode contributions,
   respectively.}
   \label{fig:2pt}
\end{figure}

The heavy-light 2-point correlator is written
\begin{equation}
C(t) = \frac{1}{V} \sum_{\vec{x},\vec{y}}
        \sum_{\vec{r},\vec{w}}
       \left\langle {\rm Tr}\left[
      S_b(\vec{y},0;\vec{x},t) \Gamma
      \bar{Q}(\vec{x}+\vec{r},t;\vec{y}+\vec{w},0) \tilde{\Gamma}_5
      \phi(\vec{r}) \phi(\vec{w})
  \right] \right\rangle,
\label{eq:correlator4}
\end{equation}
where $S_b$ is the heavy quark propagator, and $\Gamma$ and
$\tilde{\Gamma}_5$
are given in Table ~\ref{tab:gamma}.
$C(t)$ is decomposed into two parts:
\begin{equation}
  C(t) = \left\langle C_0(t) + C_1(t)\right\rangle,
\end{equation}
\begin{eqnarray}
C_0(t) &=& \frac{1}{V}
        \sum_{i=1}^{N_{\rm ev}} \frac{1}{\lambda_i}
        \sum_{\vec{x},\vec{y}} \sum_{\vec{r},\vec{w}}
      {\rm Tr}\left\{
      \gamma_5 S_b^\dag(\vec{x},t;0)
       \delta(\vec{y}-\vec{x}) \gamma_5 \Gamma
      \right. \nonumber\\
 & &  \hspace{2.8cm}\left. \times
      \left[ v^{(i)}(\vec{x}+\vec{r},t)\otimes
             v^{(i)\dag}(\vec{y}+\vec{w},0) \right]
       \tilde{\Gamma}_5
      \phi(\vec{r}) \phi(\vec{w})
  \right\},
\label{eq:correlator5a} \\
C_1(t) &=& \frac{1}{V}
         \frac{1}{N_r} \sum_{r,j}
        \sum_{\vec{x},\vec{y}} \sum_{\vec{r},\vec{w}}
      {\rm Tr}\left\{
      \gamma_5 S_b^\dag(\vec{x},t;0)
       \delta(\vec{y}-\vec{x}) \gamma_5 \Gamma
      \right. \nonumber\\
 & &  \hspace{2.8cm}\left. \times
      \left[  \psi_{[r]}^{(j)}(\vec{x}+\vec{r},t) \otimes
              \eta_{[r]}^{(j)\dag}(\vec{y}+\vec{w},0) \right]
       \tilde{\Gamma}_5
      \phi(\vec{r}) \phi(\vec{w})
  \right\}.
\label{eq:correlator5b}
\end{eqnarray}
Figure \ref{fig:2pt} schematically depicts the contributions from the low-modes and high-modes. 

While the above functions $C_0(t)$ and $C_1(t)$ are represented
with a fixed-source time slice, $t=0$, when the all-to-all
propagator technique is applied, all the translationally
equivalent correlators are averaged over.
We note that one does not necessarily average over all the
source time slices, because after the time dilution is applied,
the source time slice can be chosen arbitrarily.
The number of source time slices adopted should be chosen
appropriately for each $C_0$ and $C_1$
by considering the statistical errors and numerical
cost.
In the case of the 2-point correlation function, we average
over all the available source time slices for both the
low-mode and high-mode parts.

\begin{figure}[tb]
  \begin{center}
\vspace{-0.2cm}
      \includegraphics[width=6.0cm]{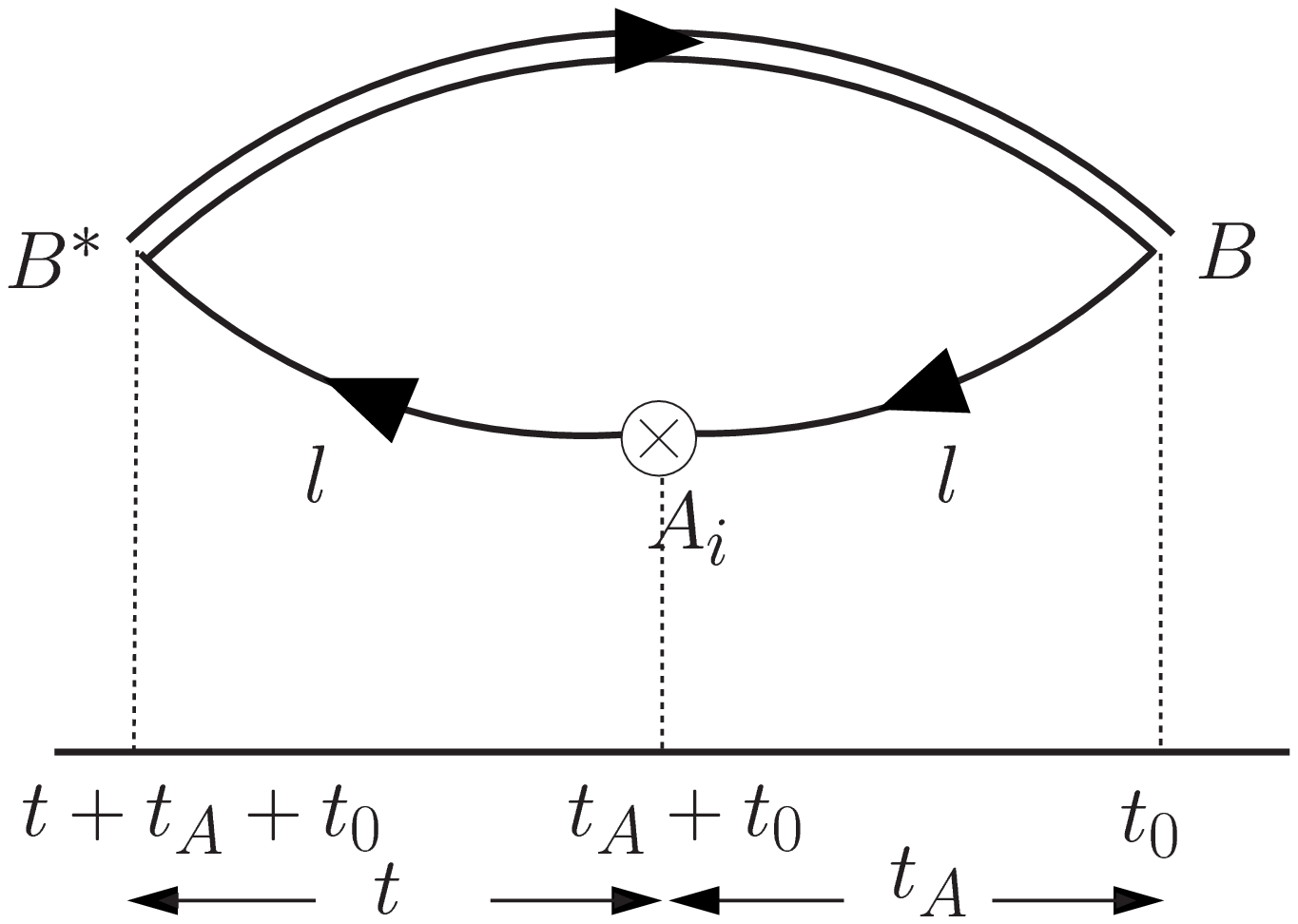}
\hspace{1cm}
      \includegraphics[width=6.cm]{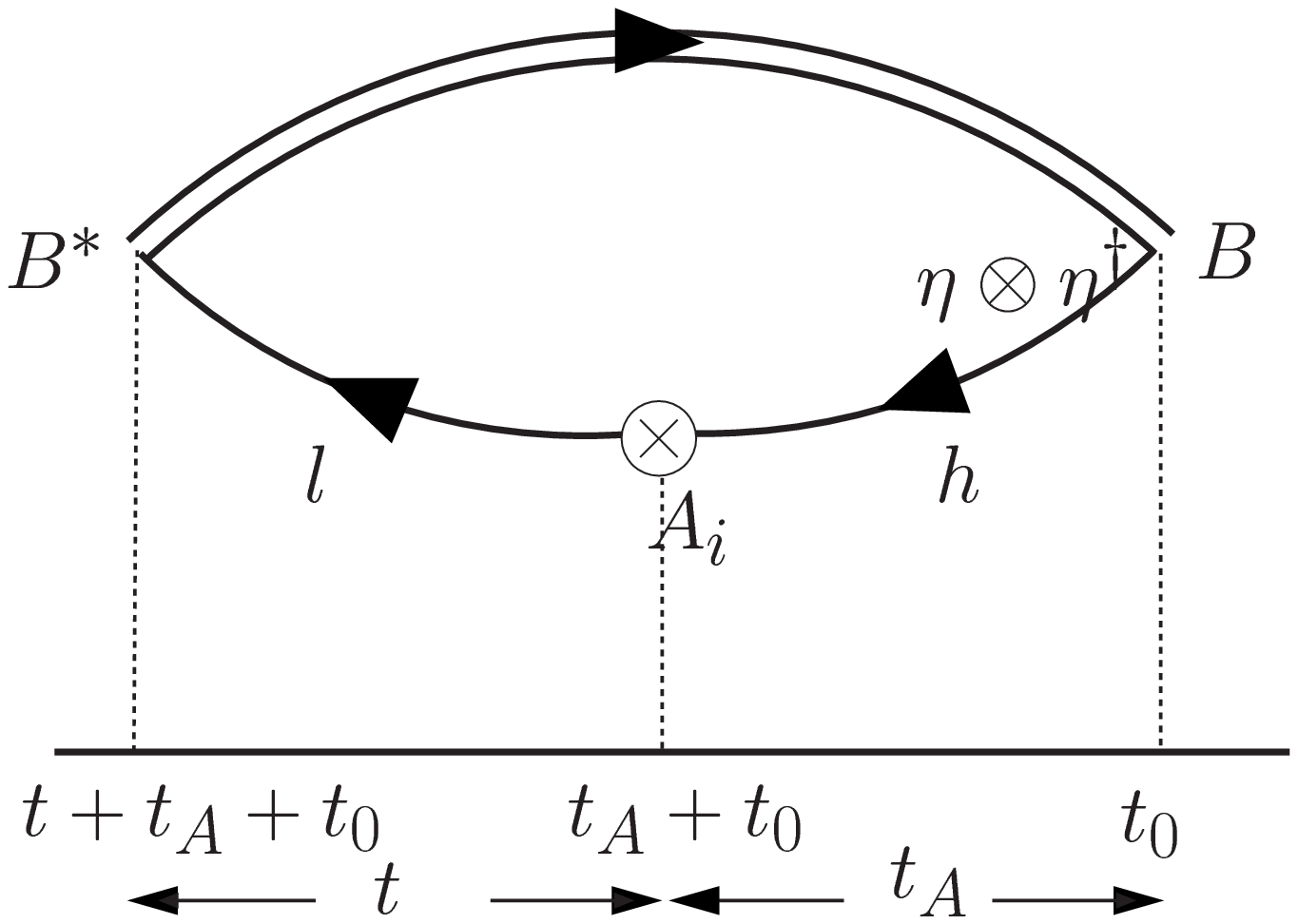}\\
\vspace{0.5cm}
      \includegraphics[width=6.cm]{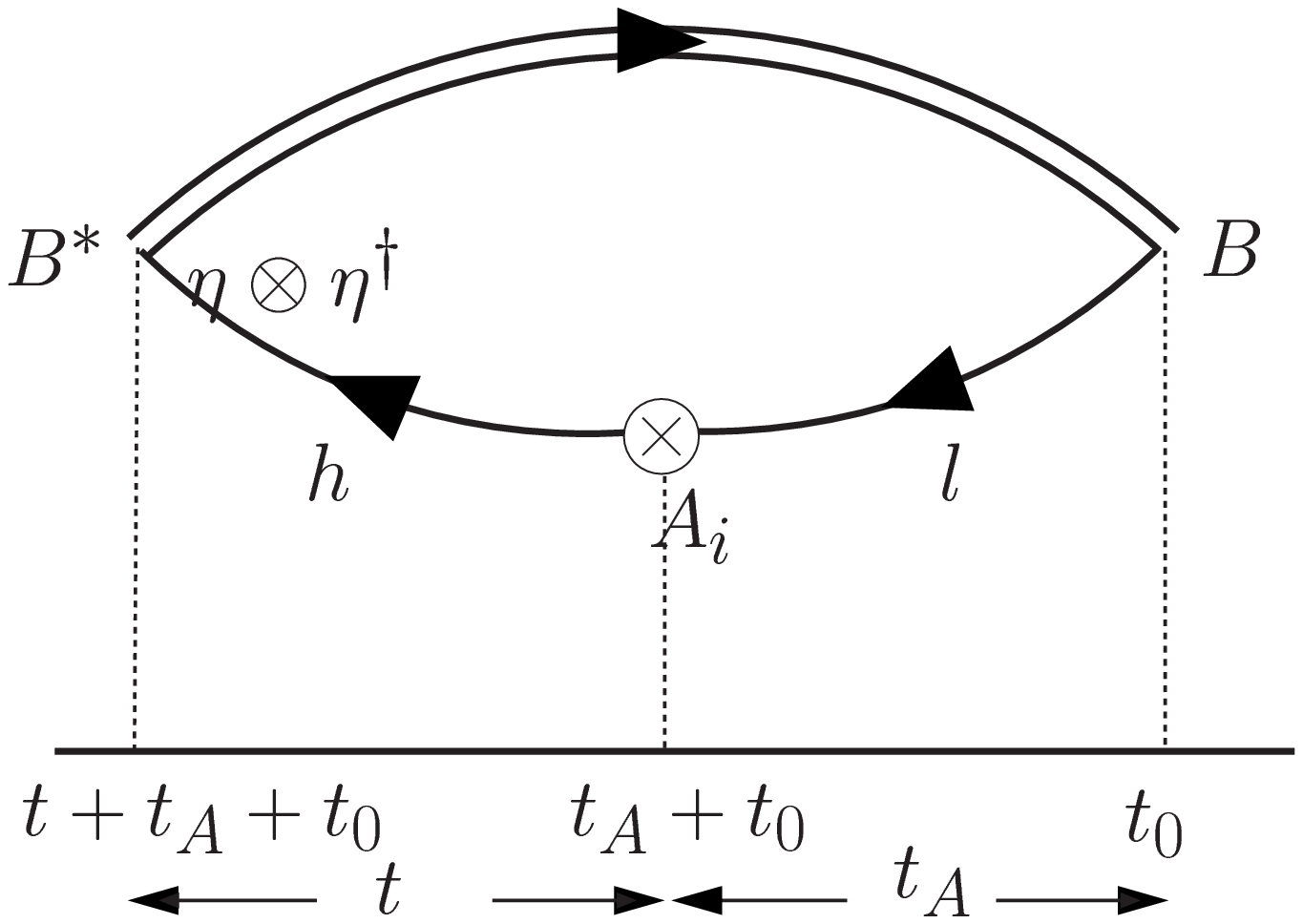}
\hspace{1cm}
      \includegraphics[width=6.cm]{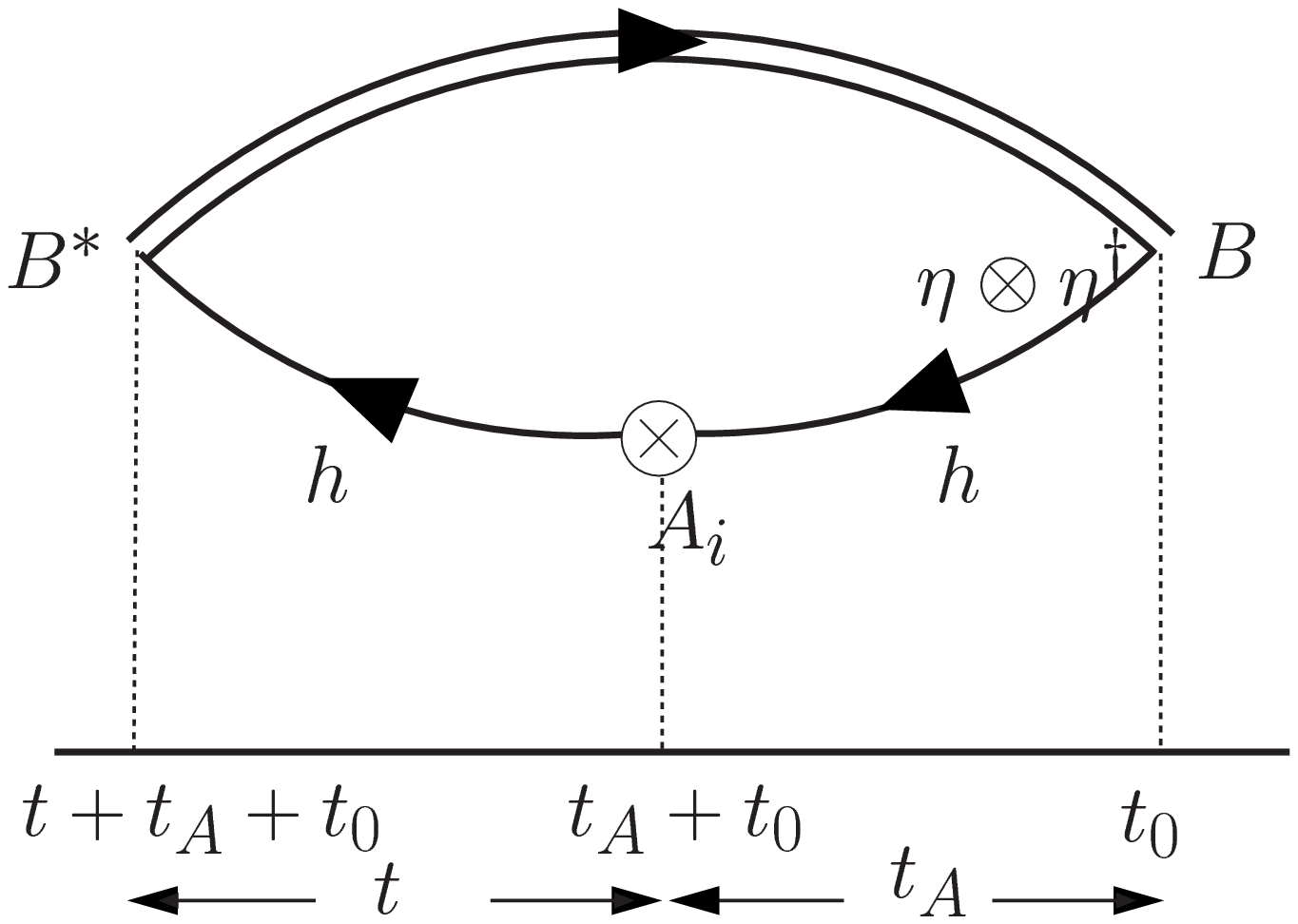}\\
%\vspace{-5cm}
\end{center}
   \caption{The diagrams for the 3-point correlation function.
The top left and right panels display the lower-lower and lower-higher parts and
bottom left and right panels display the higher-lower and higher-higher parts.}
\label{fig:3pt}
\vspace{-0.2cm}
 \end{figure}

The heavy-light 3-point correlator is given by
\begin{eqnarray}
C(t_{B^*},t_{A},t_B)
&=&\frac{1}{V}\sum_{\vec{x},\vec{y},\vec{z}}
        \sum_{\vec{r},\vec{w}}
       \Big\langle {\rm Tr}\Big[
      S_b(\vec{y},t_B;\vec{x},t_{B^*}) \Gamma_{B^*_\alpha} 
  \bar{Q}(\vec{x}+\vec{r},t_{B^*};\vec{z},t_{A})\nonumber\\
 & & \left.\left.\times \Gamma_{A_\alpha}
      \bar{Q}(\vec{z},t_{A};\vec{y}+\vec{w},t_B)
      \tilde{\Gamma}_B \phi(\vec{r}) \phi(\vec{w})
  \right] \right\rangle,
\label{eq:correlator5-1}
\end{eqnarray}
where $\Gamma_{B^*_\alpha}=\gamma_\alpha$,
$\Gamma_{A_\alpha}=\gamma_\alpha$, and $\tilde{\Gamma}_B=-1$
($\alpha=1,2,3$).
They are decomposed into four parts:
\begin{eqnarray}
  C(t_{B^*},t_{A},t_B) 
&=& \Big\langle C_{ll}(t_{B^*},t_{A},t_B) 
              + C_{lh}(t_{B^*},t_{A},t_B)
\nonumber\\
& & \hspace{0.3cm}   + C_{hh}(t_{B^*},t_{A},t_B) 
             + C_{hl}(t_{B^*},t_{A},t_B)\Big\rangle.
\end{eqnarray}
Here, $C_{ll}$, $C_{lh}$, $C_{hh}$, and $C_{hl}$ denote
the low-low, 
low-high, high-high, and high-low mode correlators,
respectively, which are schematically depicted in Fig.~\ref{fig:3pt}.
Explicit expressions of the 3-point functions are given in the Appendix.

%Now,in 2-point correlators, we have them averaged over all time slice sources(all-to-all).
%On the other hand in 3-point correlators, lower-lower parts are averaged over all time slice sources
%but other parts are taken one time slice sources.
As noted for the 2-point correlation functions,
 translationally equivalent correlators can be averaged
over an appropriate number of source time slices.
For the low-low part, we averaged over all the source time slices,
since the numerical cost is small.
For the other three parts, we explore the best solution in the following
section.

%%%%%%%%%%%%%%%%%%%%%%%%%%%%%%%%%%%%%%%%%%%%%%%%%%%%%
%
\section{Simulation details}
Our simulations were carried out on a quenched $16^3\!\times48$ lattice
with the standard plaquette gauge action $\beta=6/g^2=6.0$.
We considered 32 gauge configurations
generated by the pseudo-heat-bath update algorithm,
each separated by 1000 Monte Carlo sweeps, after
a period of 1,000 sweeps allowed for thermalization.

In the HQET, we used the static action with HYP smeared links and the
parameter value $(\alpha_1,\alpha_2,\alpha_3)=(0.75, 0.6, 0.3)$.
We used the $\cal O\it(a)$-improved Wilson fermion for the light quark
with the nonperturbatively determined clover coefficient
$c_{\rm SW}=1.769$~\cite{Luscher:1996ug}.  We considered three hopping
parameters, $\kappa=0.1335$, 0.1340, and 0.1342.
The $B$ and $B^*$ meson operators at both the source
and sink points were smeared with the smearing function
$\phi(t,\vec{r}) \propto \exp(-0.45|\vec{r}|)$.
This smearing function was
obtained from the wavefunction given in Ref.24) %\cite{Duncan:1994uq} 
by properly scaling the parameters to our lattice spacing.
The low-lying eigenmodes of $Q$ were obtained with the implicitly
restarted Lanczos algorithm~\cite{Sorensen}, accelerated by spectral
transformation with the Chebyshev polynomial of degree 80~\cite{Neff:2001zr}.
The low-mode parts of the correlation functions
can be computed with these eigenvectors.
For the high-mode parts of the correlator, we obtained the
quark propagator with a source vector which is given by
a normalized complex $Z_2$ random noise and then projected into
the space perpendicular to the space spanned by the low modes.
Then, the time, color, and spinor dilution was applied, and
all color and spinor components were summed over, while
the number of the time dilution, $N_{t_0}$, may be kept less
than the total time $N_t=48$. 
The Dirac operator was then inverted using the BiCGStab algorithm 
with the stopping condition $10^{-10}$.
The number of eigenmodes for the low-mode averaging 
was chosen to be $N_{\rm ev}=100$ and the number of the random noise and the time
dilution was taken to be $N_r=1$ and $N_{t_0}=4$ for our final
productive run. This choice was made so as to realize good
statistical accuracy of the correlators while keeping the numerical cost at
a reasonable level, based on our exploratory study of the 2-point and 3-point
functions by changing the parameter choices. This is explained
below. 
%-------------------------
\subsection{Exploratory study of the parameter choice for all-to-all propagators}
%-------------------------
In this subsection, we report the results of exploratory studies to determine the optimal parameter value
for the all-to-all propagators in order to realize maximum numerical accuracy
of the correlators with modest numerical cost. We found that
obtaining 100 low eigenmodes is feasible, because it costs about two hours per
configuration of our computational resources;~\footnote{The computational costs cited in this
section, are the CPU times using
 one CPU of our vector supercomputer. However, the CPU time depends
strongly on the architecture, and therefore a comparison of the cost
 does not have absolute meaning. Instead, if should be understood
an only one particular reference value for guidance, among many others.}
 this cost is not large compared to that of the other parts of the
calculation. For the correlator, the numerical cost for the all-to-all
propagators for the low-mode part grows linearly in $N_{\rm ev}$ for 2-point 
correlators, since there is only one light quark propagator in the
diagram, while for 3-point correlators it 
grows quadratically in $N_{\rm ev}$, because there are two light quark
propagators in the diagram. Then, the numerical cost for the high-mode part
grows linearly both in $N_{t_0}$ and $N_{r}$, because we apply the
source method to compute the high-high part of the 3-point correlation
functions. 

In the exploratory analysis described below we studied the parameter
dependence using the data with 32 configurations for both 2-point and 3-point
correlators.  All the studies of the 2-point correlator were carried out
with $N_r=1$. Except for the study of the $N_{t_0}$ dependence, the
2-point  functions were basically studied with $N_{t_0}=4$, while in
the final production run, we used $N_{t_0}=48$.  For the 3-point
correlators, basic studies were carried out for $(N_r,N_{t_0},t_A)=(1,4,8)$
except for the study of the $(N_r,N_{t_0})$ and $t_A$ dependences, while for
the production run, the same parameter values were adopted. 
\subsubsection{$N_{\rm ev}$ dependence of the 2-point correlator}
\begin{figure}[tbp]
\begin{center}
\vspace{-1.0cm}
\includegraphics[width=9cm]{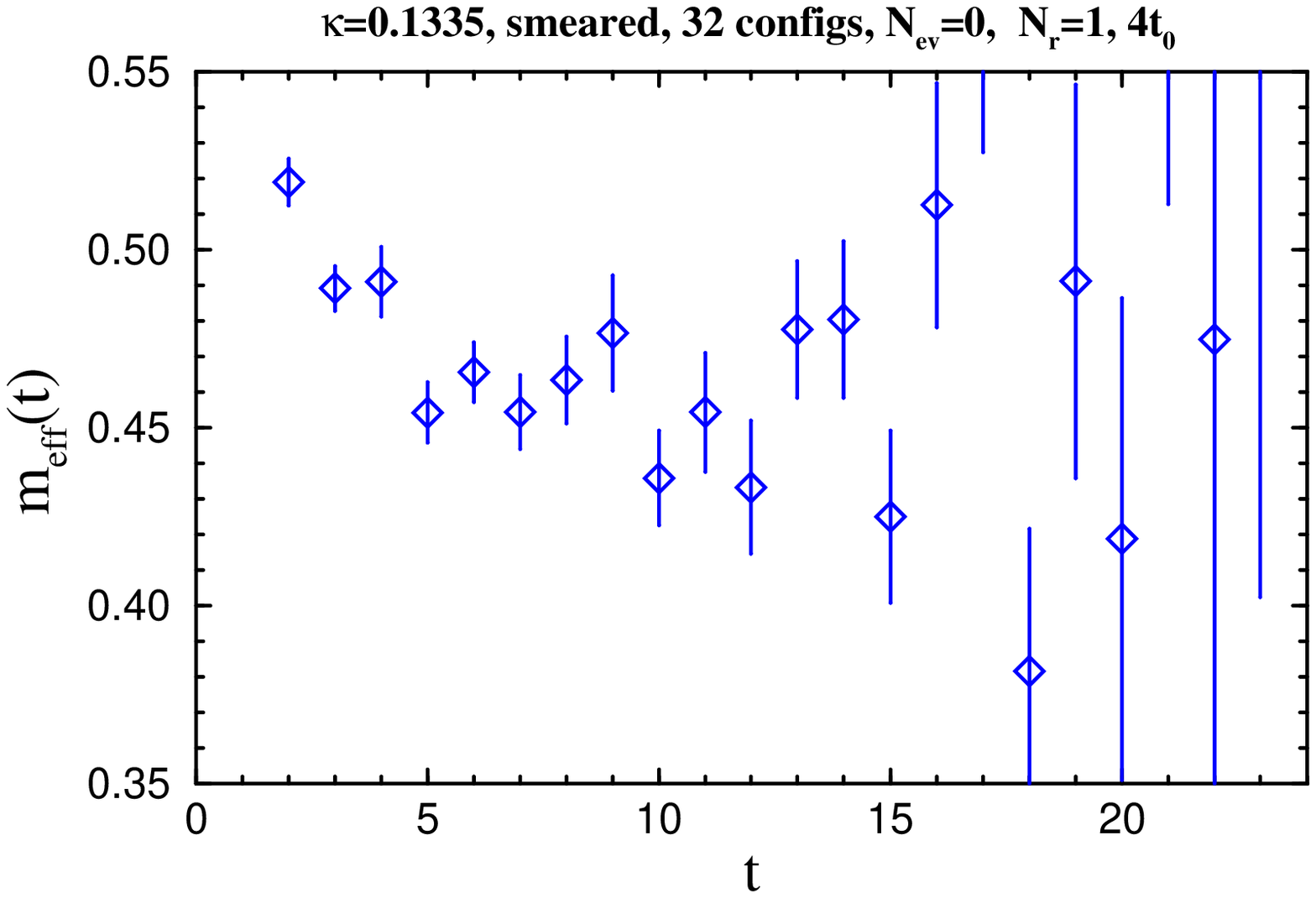}\\
\vspace{-2.2cm}
\includegraphics[width=9cm]{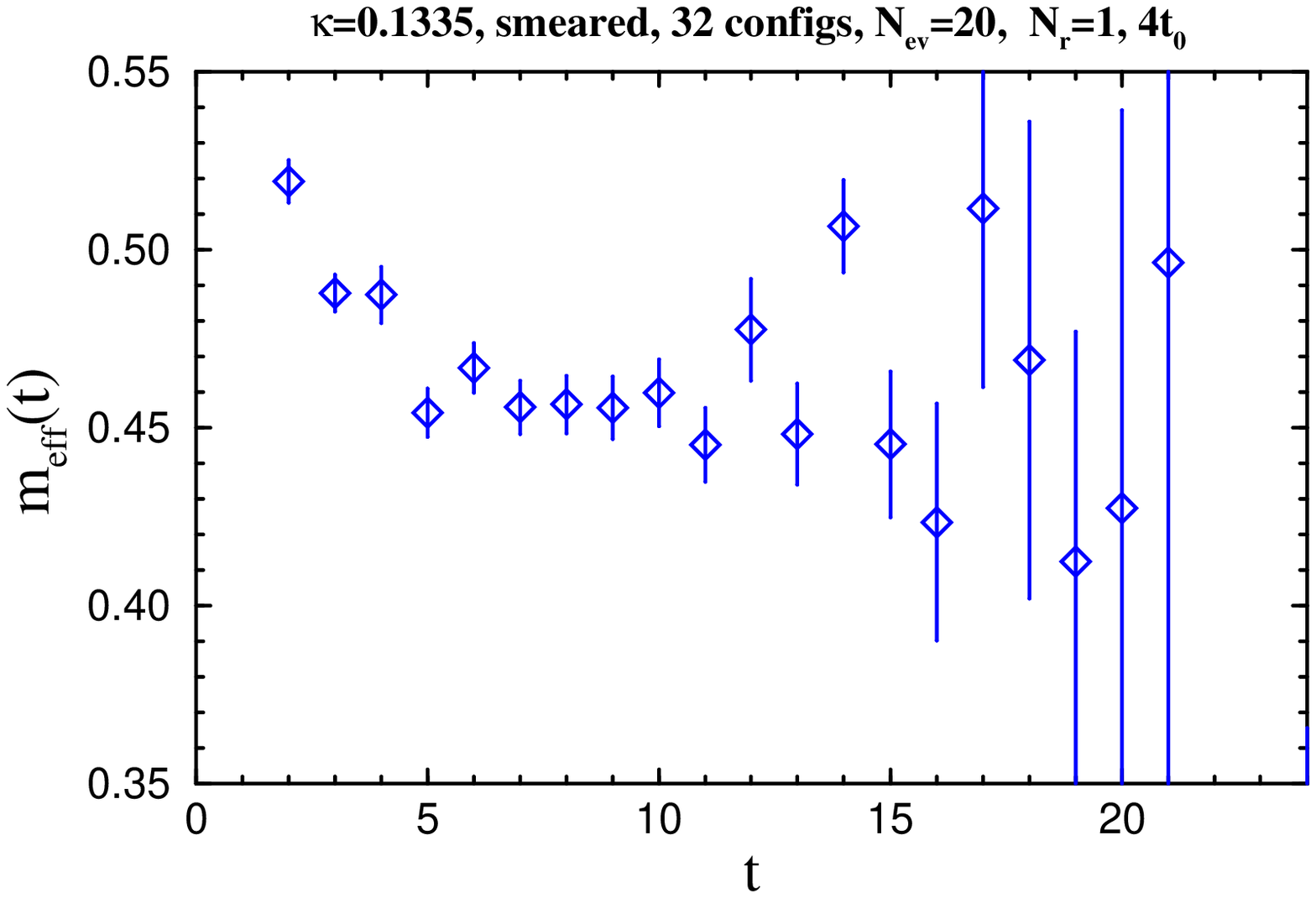}\\
\vspace{-2.2cm}
\includegraphics[width=9cm]{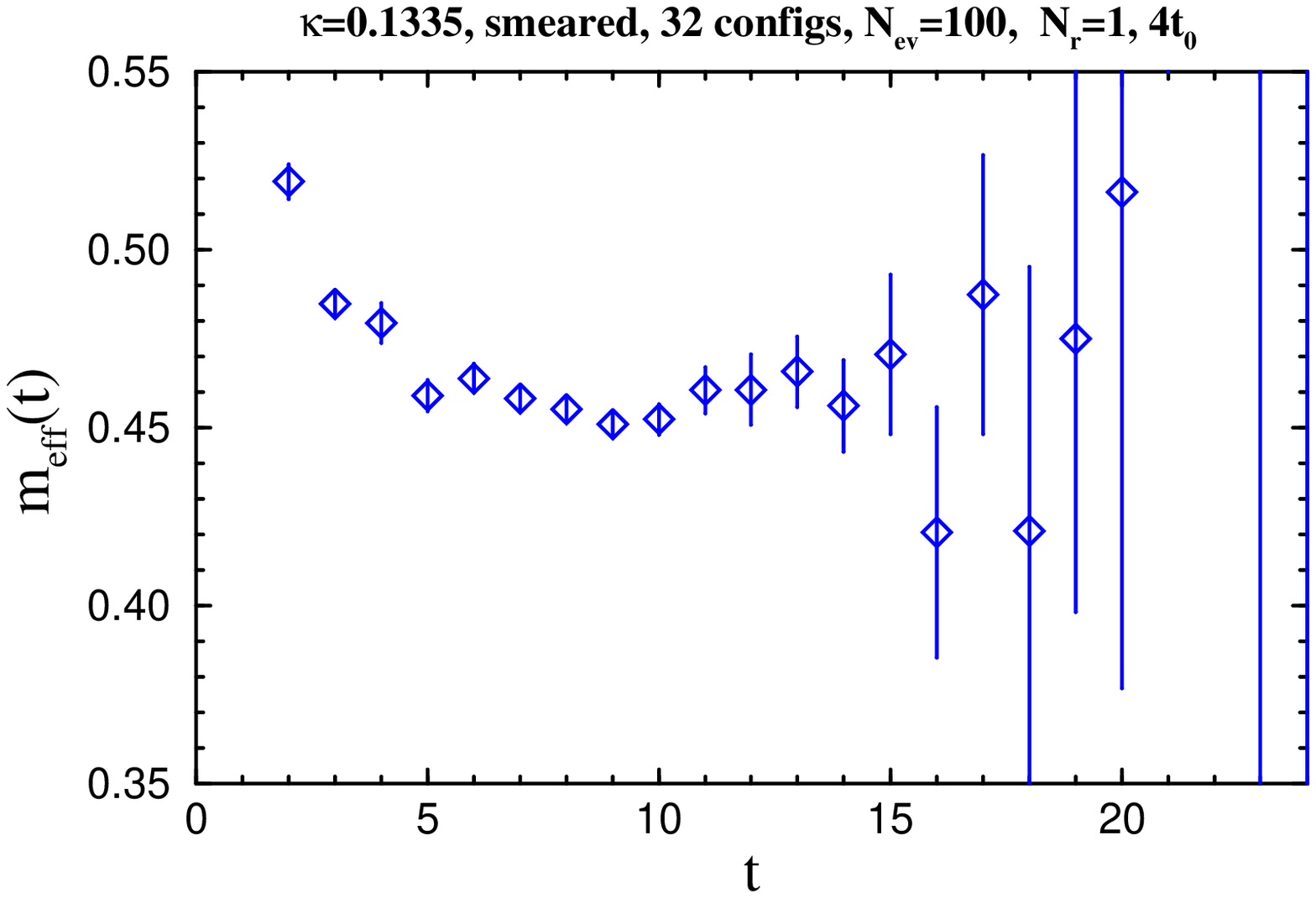}\\
\vspace{-1.3cm}
  \end{center}
\caption{Plots of the effective mass of the 2-point
 correlators for $\kappa=0.1335$, with low mode averaging and $N_{t_0}=4$
 and $N_r=1$ for high modes. The top, middle and bottom panels correspond
 to $N_{\rm ev}=0$, 20, and 100, respectively.}
\label{fig:Nev_dep_2pt_meff_k1335}
\vspace{-0.2cm}
\end{figure}
We first study the $N_{\rm ev}$ dependence of the 2-point correlators. 
The low-mode averaging gives the exact all-to-all propagator for the low
mode part, and thus it improves the statistical accuracy of the low-mode
contribution significantly. Therefore if the low-mode contribution
dominates the 2-point function for the time range where the effective
mass exhibits a plateau, one can expect that better accuracy can be
obtained.  
Figure~\ref{fig:Nev_dep_2pt_meff_k1335} plots the effective mass 
of the 2-point correlators for three choices of the 
number of low eigenmodes, $N_{\rm ev}=0, 20$ and $100$, with $N_r=1$ for
$\kappa=0.1335$. The statistical fluctuations apparently 
 decrease as $N_{\rm ev}$ increases, and for $N_{\rm ev}=100$ we 
observe a clear plateau for $t \geq 7$. 
The situation is almost the same for $\kappa=0.1342$.
\begin{figure}[tbp]
\begin{center}
\vspace{-1.0cm}
\includegraphics[width=9cm]{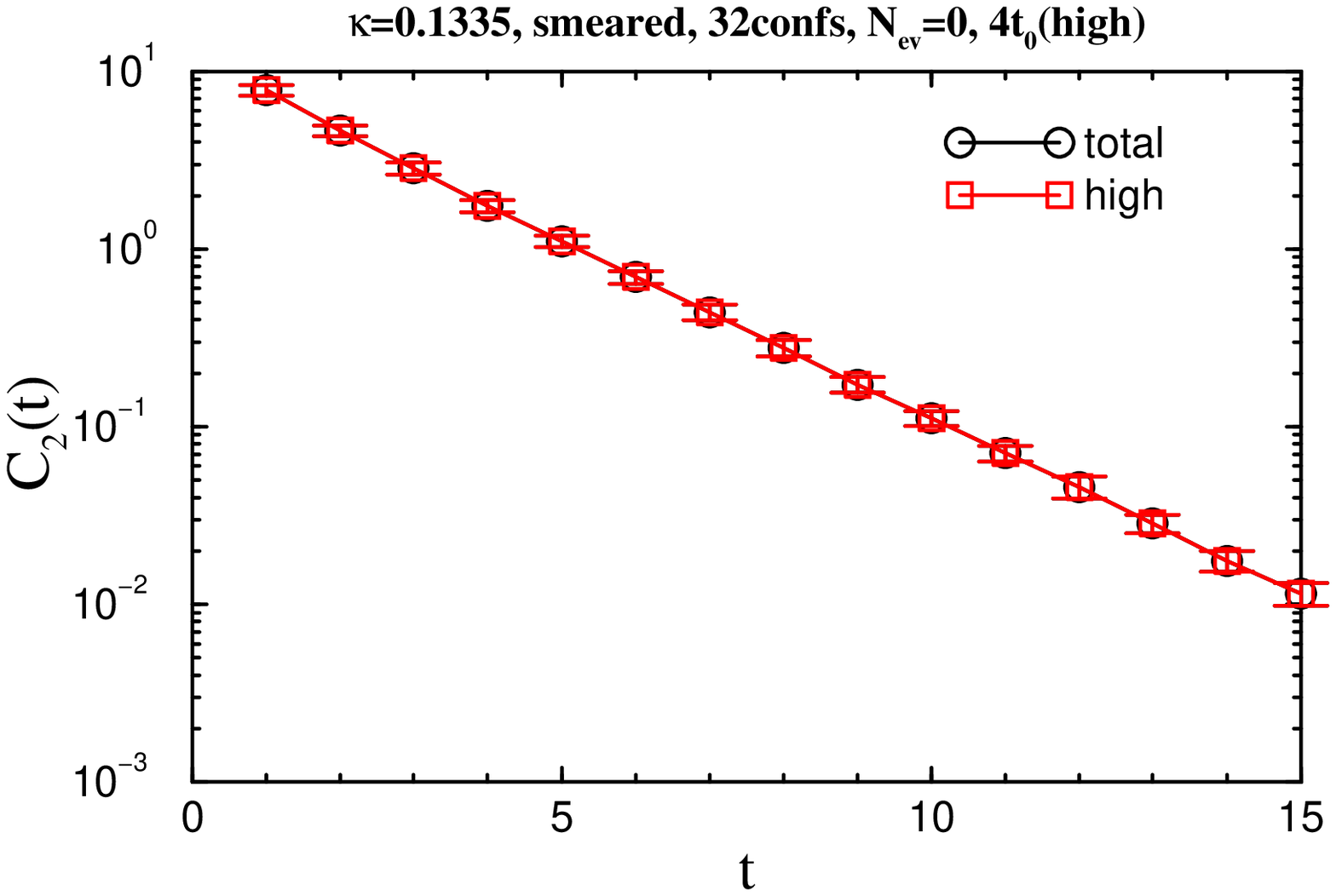}\\
\vspace{-2.2cm}
\includegraphics[width=9cm]{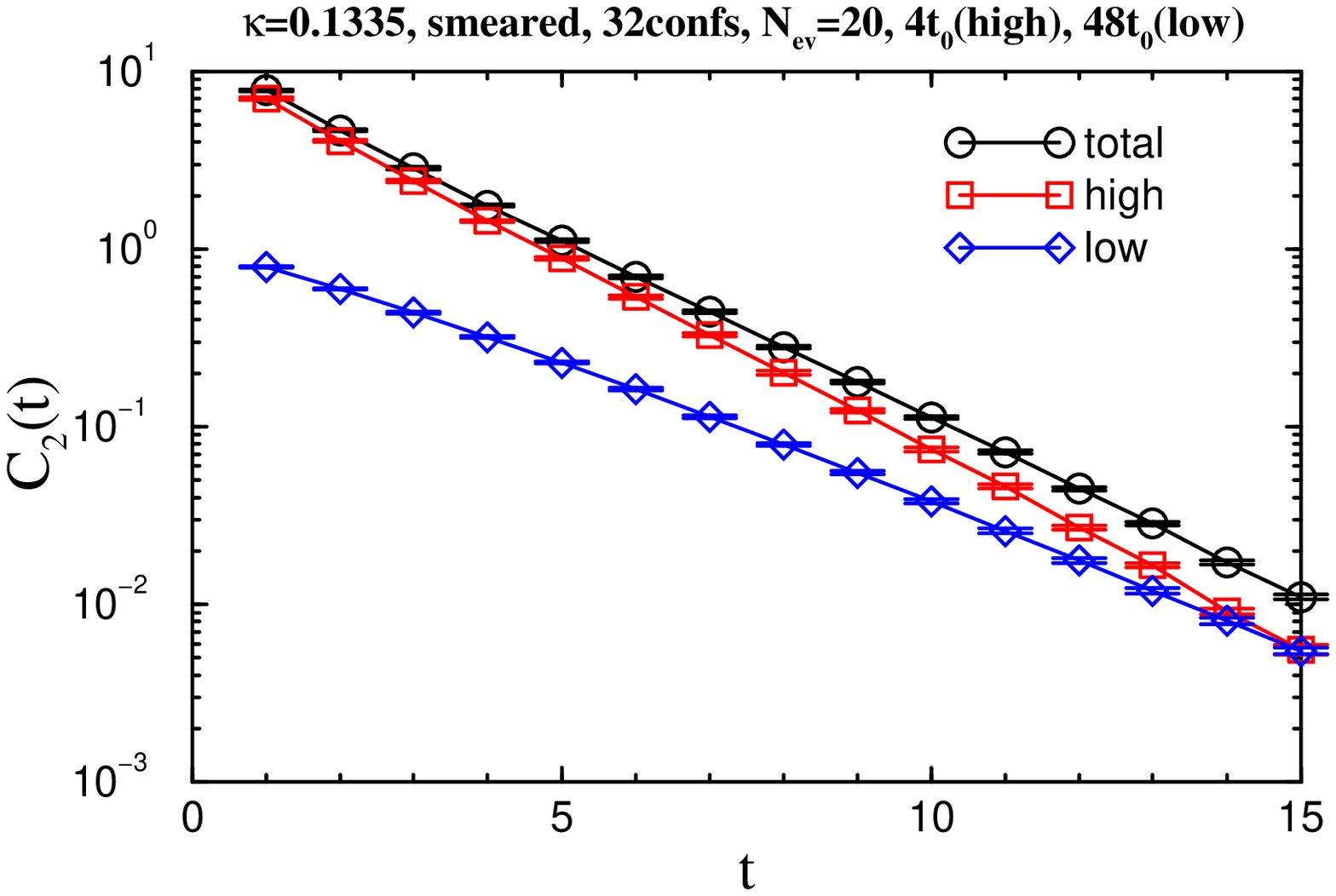}\\
\vspace{-2.2cm}
\includegraphics[width=9cm]{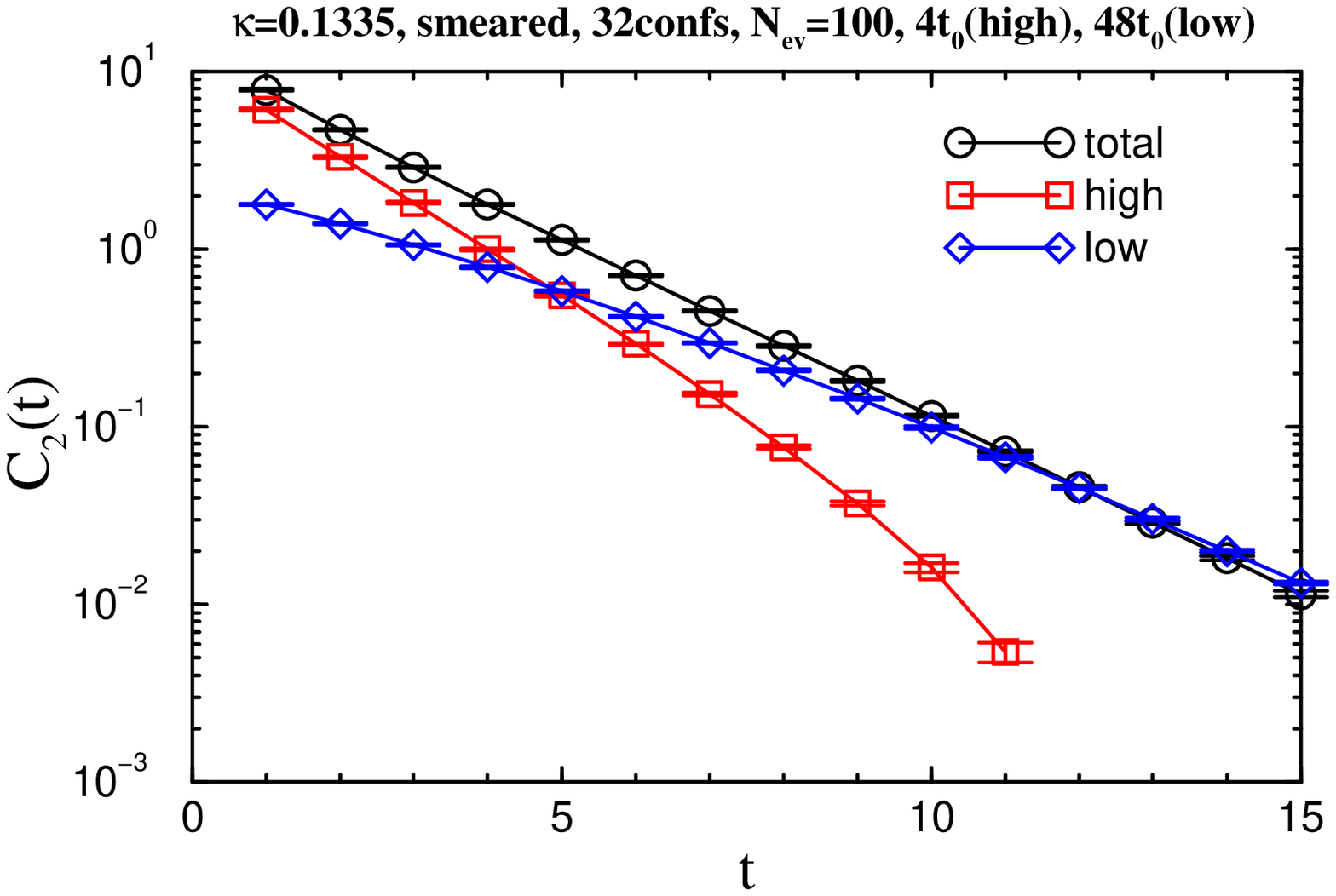}
\vspace{-1.3cm}
  \end{center}
   \caption{Low-mode and high-mode contributions to the 2-point
 correlators for $\kappa=0.1335$ with low-mode averaging and $N_{t_0}=4$
 and $N_r=1$ for the high modes. The top, middle and bottom panels correspond to
$N_{\rm ev}=0$,20, and 100, respectively.}
\label{fig:Nev_dep_2pt_k1335}
\vspace{-0.2cm}
\end{figure}

The behavior described above can be understood as the low-mode dominance in the correlator for
large $t$. Because we used the low-mode averaging, the low-mode part 
has the best accuracy.
For the high-mode part, we only averaged over
$N_{t_0}=4$ timeslices for source points in this particular analysis, 
and for this reason, maximal statistical accuracy was not obtained. Also, the 
high-mode part suffers from additional statistical error due to the random 
source for the noisy estimator. These makes the high-mode part much
noisier than the low-mode part.  Figure.~\ref{fig:Nev_dep_2pt_k1335} shows
the low-mode and high-mode contributions to the 2-point correlators
for different choices of the  number of low eigenmodes, $N_{\rm ev}$. We
find that for $N_{\rm ev}=100$, the low-mode contribution dominates the
2-point correlator for $t \geq 6 $, where
the plateau has almost been reached, while for $N_{\rm ev}=20$, low-mode
dominance is obtained only for $t>15$. 

Although the 100 eigenmode calculation requiring 2 hours is rather time
consuming, it still worthwhile to consider larger computational times for 
the 3-point correlators.  Carrying out the complete
low-mode averaging of the 2-point correlators
takes about 35 minutes per configuration. The computation for the high modes with
averaging over four time slices typically takes another 35 minutes, while
averaging over all 48  time slices  would typically take 7
hours. This means that choosing $N_{\rm ev}=100$ and reducing the effect of
the noisy high mode contribution is the optimal choice.

\subsubsection{Effect of low mode averaging for the 2-point correlator} 
\begin{figure}[tb]
\begin{center}
\vspace{-1.2cm}
\includegraphics[width=10cm]
{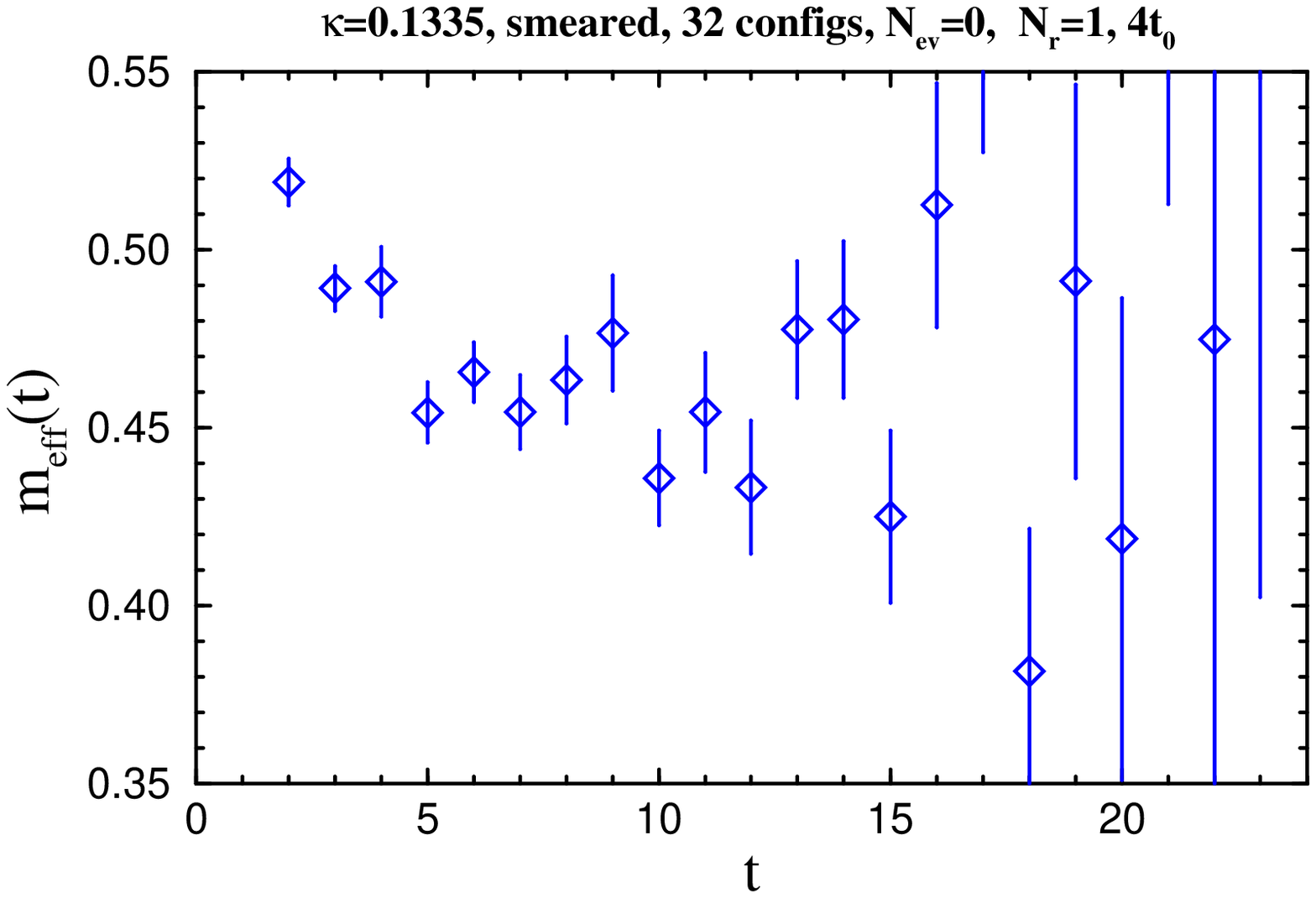}\\
\vspace{-2.2cm}
\includegraphics[width=10cm]
{./graph/Tsukuba_final/2pt/2pt_meff_k1335_Nev=100_4t0_lma.eps}
\vspace{-1.5cm}
  \end{center}
\caption{Plots of the effective mass of the 2-point
 correlators for $\kappa=0.1335$ with low mode averaging over 
4 time slices (top panel) and 48 time slices (bottom panel), with  $N_r=1$ 
and  $N_{\rm ev}=100$.}
\label{fig: low_timesilce_dep_2pt_meff_k1335}
\vspace{-0.2cm}
\end{figure}
Although it is obvious that once we have the low eigenmodes 
we should carry out complete low-mode averaging,  
 it would still be interesting to see how much gain in statistical
accuracy we obtain with the low-mode averaging. For this purpose, we
compare the results obtained with averaging over only four source time slices  and
those obtained with averaging over all time slices for the low-mode part. 
Figure~\ref{fig: low_timesilce_dep_2pt_meff_k1335} presents a comparison 
of the effective mass plots of the 2-point correlators. There, the 
low-mode contributions were averaged over four equally time slices
separated and over 48 timeslices. There is obviously a significant
improvement in the statistics. Considering the errors for $t=$5--10 we find
that increasing $N_{t_0}$ from 4 to 48, the statistical error is
further reduced by a factor of 2--3, which is effectively  equivalent to
having 4-9 more statistically independent configurations. This
suggests that out of the 48 time slices of 8--12 time slices are effectively
statistically independent.  
\subsubsection{Effect of $t_0$ averaging for the high-mode part  for the 
2-point correlator}
\begin{figure}[tb]
\begin{center}
\vspace{-1.2cm}
\includegraphics[width=10cm]
{./graph/Tsukuba_final/2pt/2pt_meff_k1335_Nev=100_4t0_lma.eps}\\
\vspace{-2.2cm}
\includegraphics[width=10cm]
{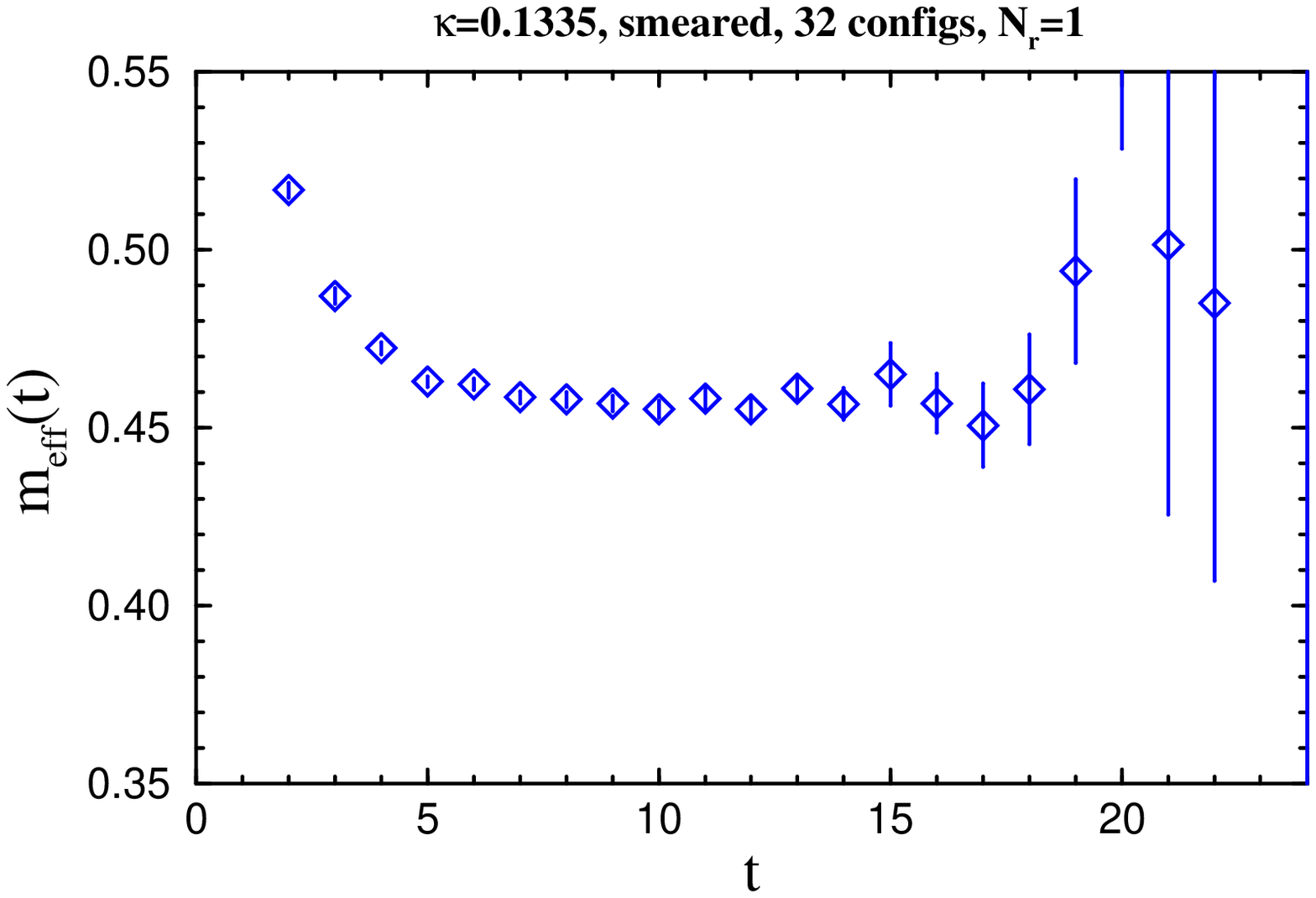}
\vspace{-1.5cm}
  \end{center}
\caption{Plots of the effective mass of the 2-point
 correlators for $\kappa=0.1335$ with complete low mode averaging.
The high-mode part is averaged over 
4 time slices (top) and 48 time slices (bottom),
with  $N_r=1$ and  $N_{\rm ev}=100$.} 
\label{fig: high_timesilce_dep_2pt_meff_k1335}
\vspace{-0.2cm}
\end{figure}
We also studied the effect of $t_0$ averaging for the high-mode contribution.
In Fig.~\ref{fig: high_timesilce_dep_2pt_meff_k1335}, we compare the
data with $N_{t_0}=4$ and $N_{t_0}=48$. It is seem that we obtain a clearer
plateau for the data with $N_{t_0}=48$.  In fact, the errors for the data with
$N_{t_0}=4$ are larger by a factor of approximately 1.5--2 than those with
$N_{t_0}=48$ due to the reduction in the error for the  high mode
part, which is not a negligible contribution. This is because, even though
the high-mode contribution itself is small, its error is roughly 
equal to that of the low- mode parts. 

The data with $N_{t_0}=48$ in Fig.~\ref{fig:
high_timesilce_dep_2pt_meff_k1335} also exhibits a clean plateau which
starts at $t = 7$, or perhaps even a smaller value.  From this result,  in the
 study of 3-point correlators reported below, we chose $t_A=7$ , 8, and 9.  
\subsubsection{$N_r$ dependence of the 2-point correlator}
It is also possible to reduce the error by increasing $N_r$. However,
as pointed out by the TrinLat group, increasing $N_r$ only reduces the
error from the noisy estimator in the high mode, while 
increasing $N_{t_0}$ reduces the error from both the noisy estimator
and the fluctuation of the gauge field. Since the
numerical cost of computing the 2-point correlator is relatively small, we decided to use $N_{t_0}=48$ and
$N_r=1$, which already gives sufficiently good results. 
We postpone the actual study of the $N_r$ dependence to a later
subsection, where we study the 3-point correlators.

\subsubsection{Low mode dominance in the 3-point correlator}
 \begin{figure}[tb]
  \begin{center}
\vspace{-1.2cm}
      \includegraphics[width=11cm]
{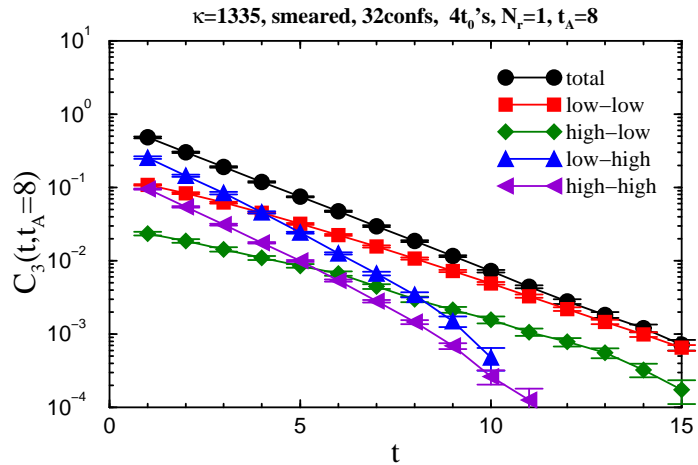}
\vspace{-1.8cm}
  \end{center}
   \caption{Contributions of the low-low, high-low, low-high and high-high 
  parts to the total 3-point function.}
     \label{fig:3pt_parts}
\vspace{-0.2cm}
 \end{figure}
Now we turn to the 3-point correlators, for which we study
the contribution of low-low, high-low, low-high and high-high modes.
Figure~\ref{fig:3pt_parts} shows the contribution of each mode to the
3-point correlator for $\kappa=0.1335$, $N_{\rm ev}=100$,
$(N_r,N_{t_0})=(1,4)$ and $t_A=8$. We find that for $t < 9$ $( t+t_A <
  17 )$ high-low and low-low are the two dominant contributions, with
high-low larger for $t\leq 4$ $(t+t_A \leq 12)$ 
and low-low larger for $t\geq 5$ $(t+t_A \geq 13)$. 

While it is true that low-mode averaging does help to reduce the
error, it is also important to reduce the error for the high-low
contribution by carrying out more averaging for the high-modes.
This is realized in two independent ways. One is to take a larger
$N_{t_0}$ for averaging the high-mode over the timeslices of the
source points. The other is to increase $N_r$. In the next subsections 
we examine the effects of both.
\subsubsection{$(N_r,N_{t_0})$ dependences of the 3-point correlator}
\begin{figure}[tb]
\begin{center}
\vspace{-1.0cm}
\includegraphics[width=10cm]
{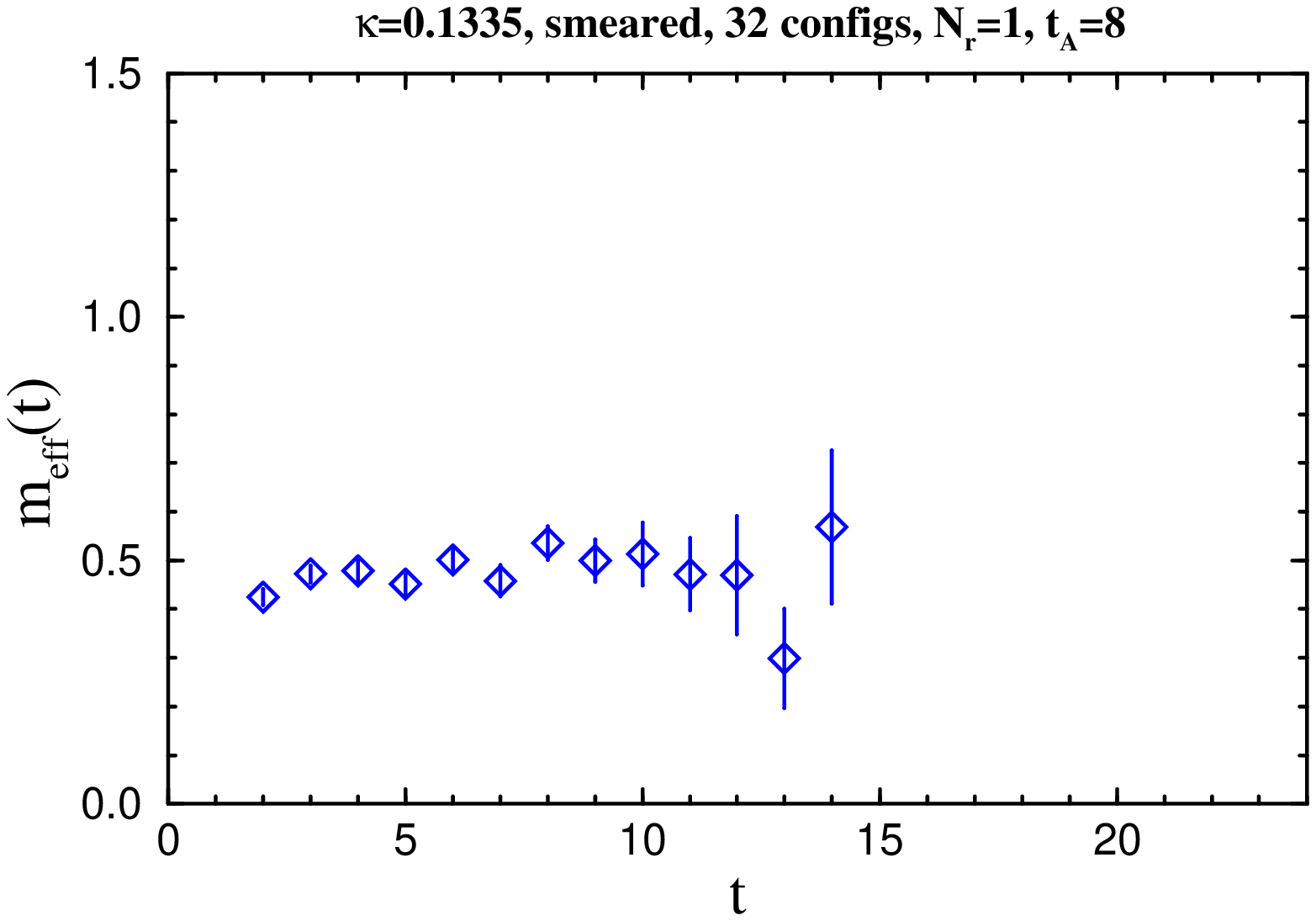}\\
\vspace{-2.2cm}
\includegraphics[width=10cm]
{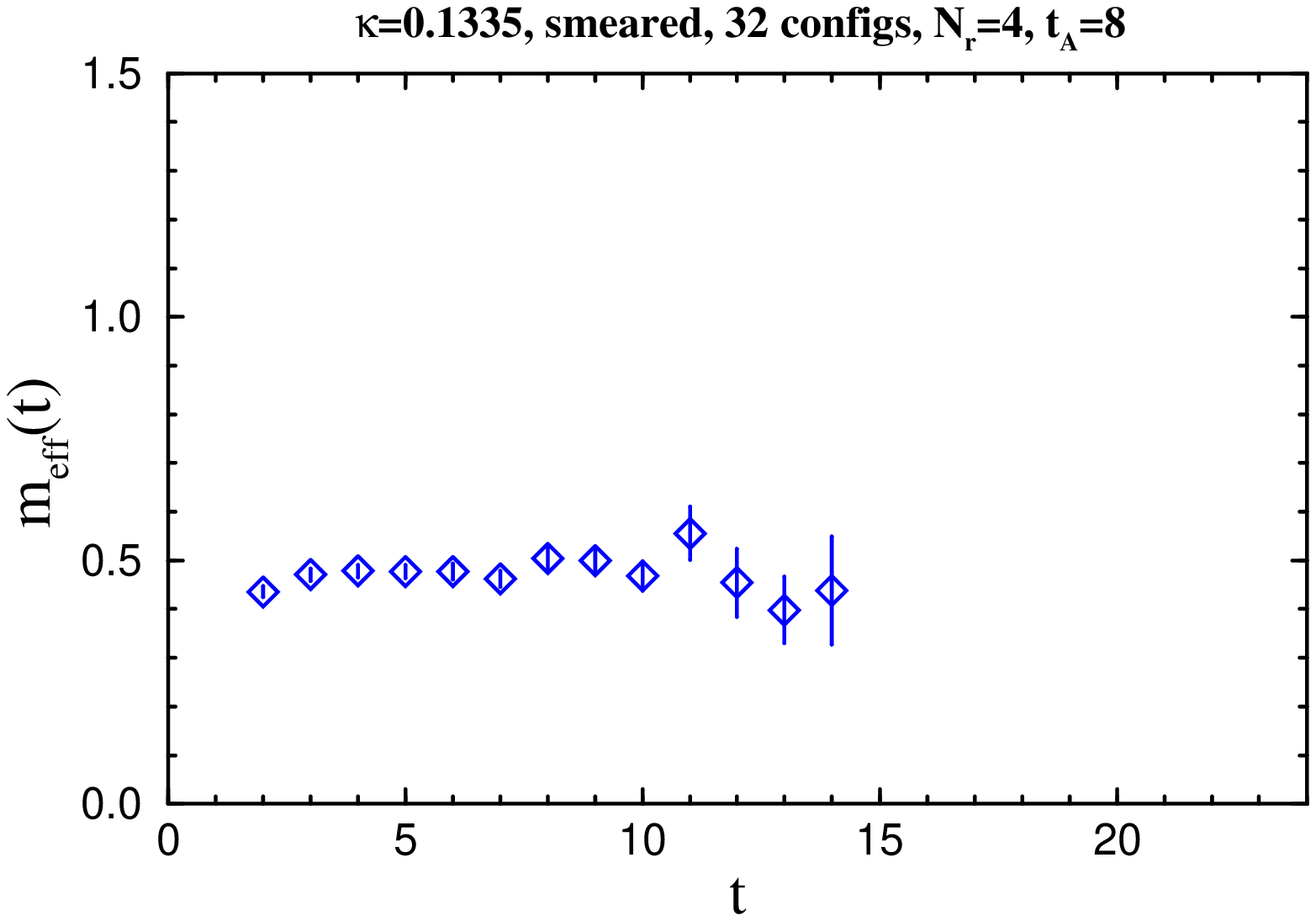}\\
\vspace{-2.2cm}
\includegraphics[width=10cm]
{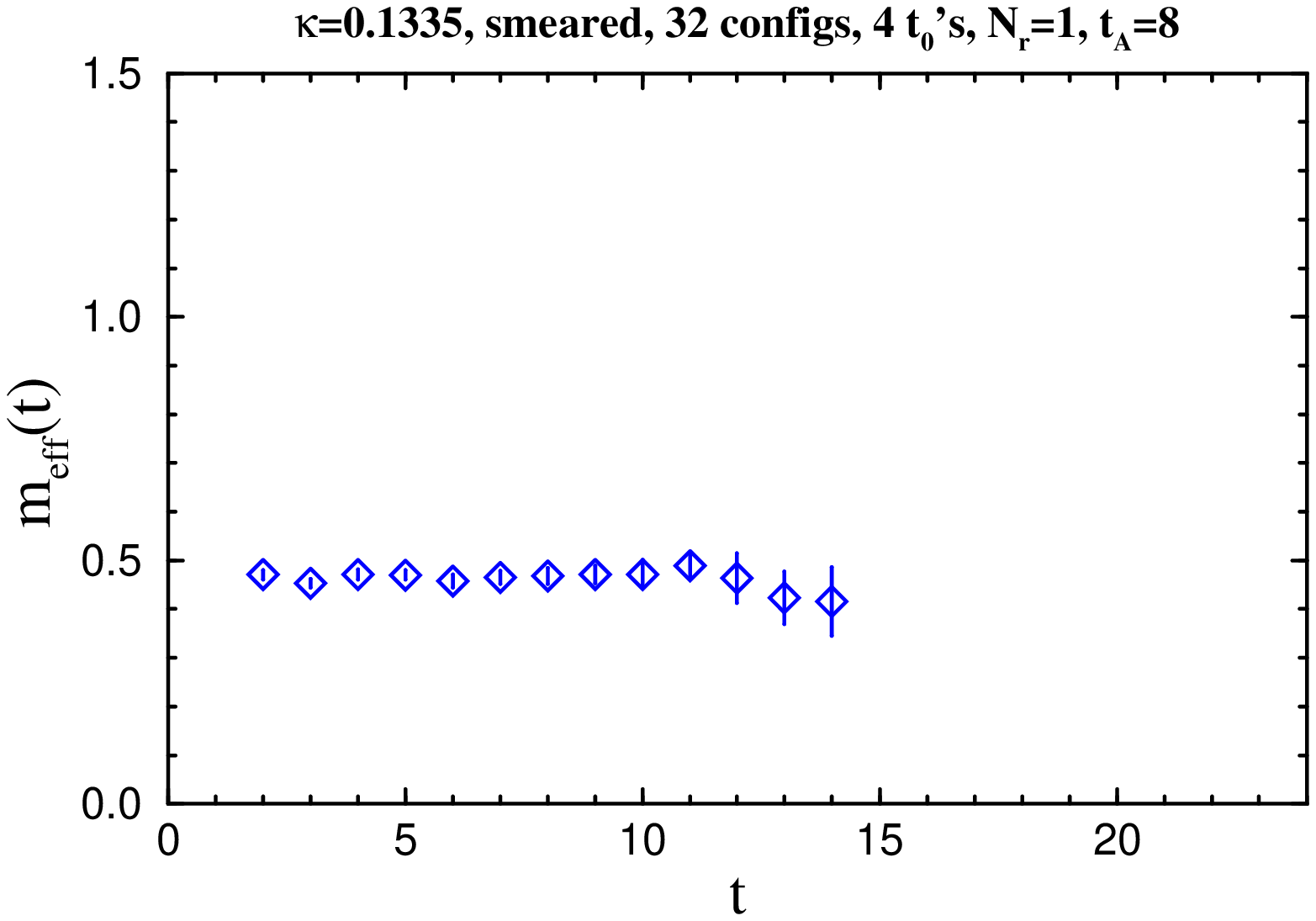}
\vspace{-1.5cm}
  \end{center}
\caption{Plots of the effective mass of the 3-point
 correlators for $\kappa=0.1335$ for $(N_r,N_{t_0})=(1,1)$(top),
$(N_r,N_{t_0})$ = (4,1) (middle) and  (1,4) (bottom), 
with $t_A=8$, $N_{\rm ev}=100$.} 
\label{fig: Nr_Nt0_dep_3pt_meff_k1335}
\vspace{-0.2cm}
\end{figure}
Figure~\ref{fig: Nr_Nt0_dep_3pt_meff_k1335} plots the $(N_r,N_{t_0})$ 
dependences of the 3-point correlator with $t_A=8$. We find that
increasing $N_r$ 
from 1 to 4 reduces the statistical errors by a factor of approximately 0.7, 
while increasing $N_{t_0}$  from 1 to 4 reduces
 the statistical errors for all $t$ by a factor of approximately 0.5. Therefore, we find that for given
computational resources, increasing $N_{t_0}$ is more efficient 
than increasing $N_r$. This is as expected, since increasing $N_{t_0}$ 
reduces both the error from the gauge fluctuation and the error from the 
noisy estimator, while increasing $N_r$ reduces only the latter.  
\subsubsection{$t_A$ dependence}
\begin{figure}[tb]
\begin{center}
\vspace{-1.2cm}
\includegraphics[width=11cm]
{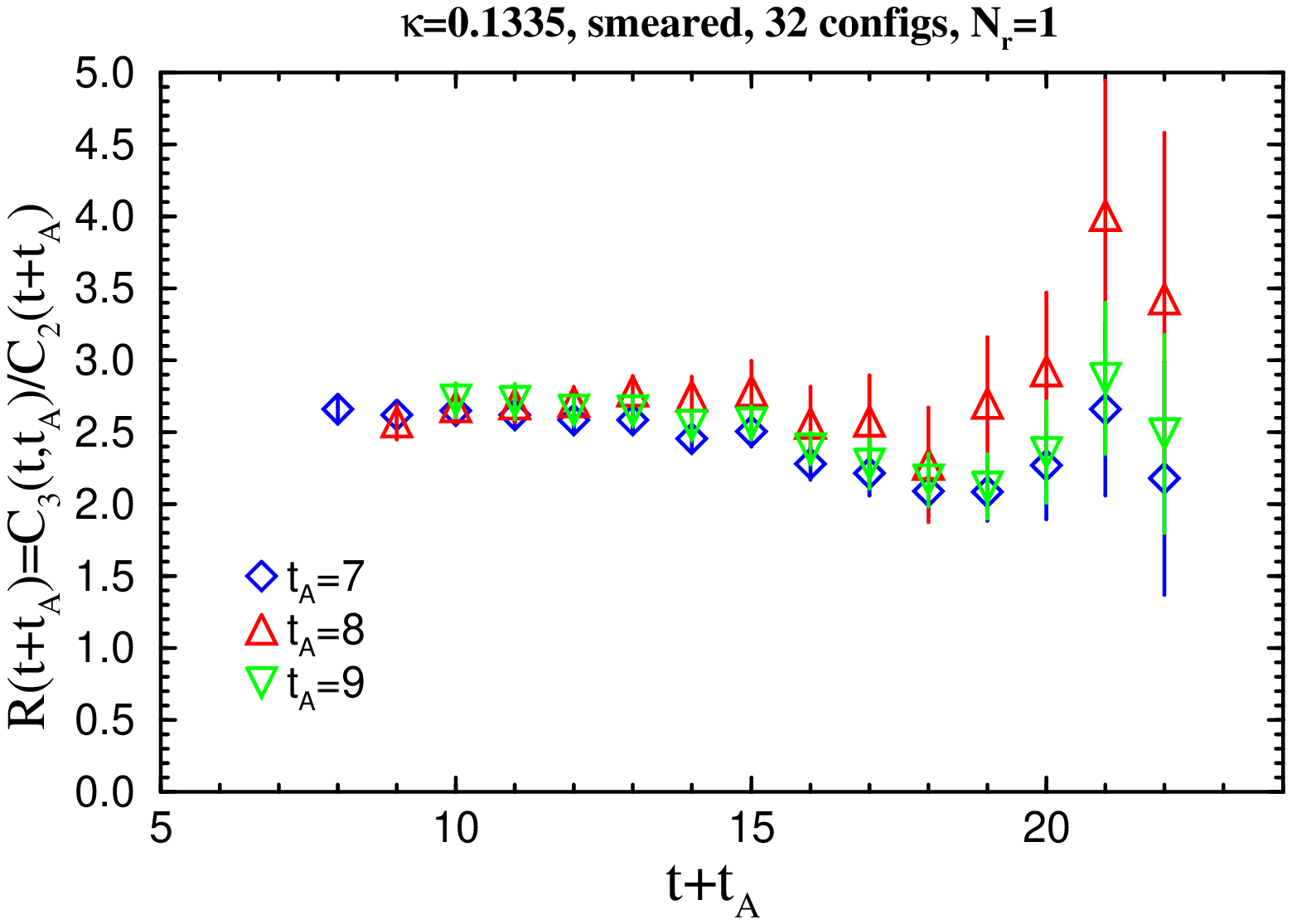}
\vspace{-1.5cm}
  \end{center}
\caption{$t_A$ dependence of the  3-point/2-point correlator ratio 
for $\kappa=0.1335$ with $(N_r,N_{t_0})=(1,1)$ 
and $N_{\rm ev}=100$.} 
\label{fig: t_A_dep_3pt_2pt_k1335}
%\vspace{-0.2cm}
\end{figure}
Figure~\ref{fig: t_A_dep_3pt_2pt_k1335} plots the time dependence of 
the 3-point/2-point ratio $ R(t,t_A)$ for $t_A=7,8$ and $9$.
The results for $t_A=7,8$ and $9$ are consistent.
The result of the data fit for $t_A=9$ 
is relatively poor and has large statistical error, while those for
$t_A=7$ and $8$ do not differ in precision. From this study, we conclude
that $t_A= 8$ is a reasonable choice for the productive run.  We also find that
the best fitting range of $R(t,t_A)$ is $t+t_A=12-16(t=4-8)$ .  
%
%%%%%%%%%%%%%%%%%%%%%%%%%%%%%%%%%%%%%%%%%%%%%%%%%%%%%
%
\section{Results}
%
%----------------------------------------------------
In this section, we present the results from our productive run for 32
cofingurations on $16^3\times 48$ lattices with $\beta=6.0$ using the $O(a)$-improved Wilson fermion with $c_{\rm SW}=1.769$. We chose three values of the
hopping parameters, $\kappa=0.1335$, 0.1340, and 0.1342. 
For the all-to-all propagators, our choices of the parameters were 
\begin{eqnarray}
N_{\rm ev}=100,\ & N_{t_0}=48,\ & N_r=1
\end{eqnarray}
for 2-point correlator, and
\begin{eqnarray}
N_{\rm ev}=100,\ & N_{t_0}=4,\ & N_r=1,\ t_A=8 
\end{eqnarray}
for 3-point correlator.
%-----------------
\subsection{2-point correlators}	
%----------------
%
\begin{figure}[tb]
  \begin{center}
\vspace{-0.8cm}
\hspace{0.5cm}
\includegraphics[width=9.5cm]
{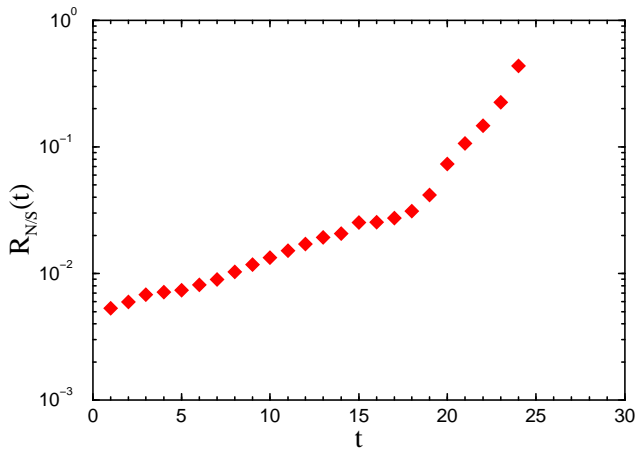} 
\vspace{-1.0cm}
  \end{center}
   \caption{The ratio $R_{NS}$ for the 2-point correlation function
 with 32 configurations for $\kappa=0.1335$.}
   \label{fig:Noise_to_signal}
\vspace{-0.2cm}
 \end{figure}
\begin{figure}[tb]
  \begin{center}
\vspace{-1.4cm}
\hspace{0.2cm}
      \includegraphics[width=11cm]{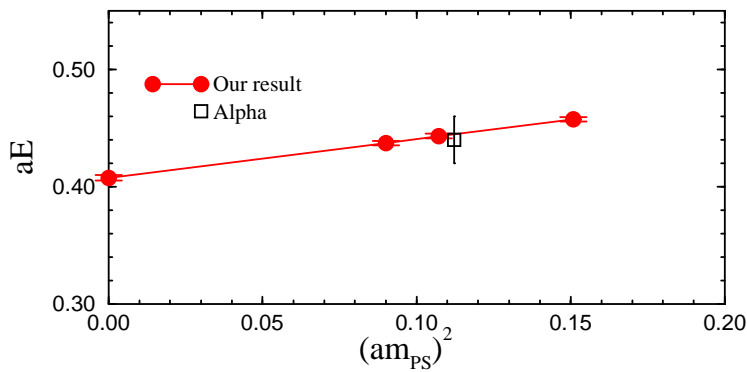}
\vspace{-2.0cm}
  \end{center}
    \caption{Binding energies from 2-point correlators. 
 The solid circles represent our data for $\kappa=0.1335$,  0.1340, and 0.1342 with
 32 configurations. The square represents the result from the Alpha
  collaboration with a few hundred configurations.}
   \label{fig:aE}
\vspace{-0.2cm}
 \end{figure}

Figure~\ref{fig:Noise_to_signal} displays the noise-to-signal ratio
$R_{NS}=\Delta C_2(t)/C_2(t)$ for the 2-point function. As shown in 
this figure the noise-to-signal ratio remains roughly
in the range 1--3 \% for the time range 1--2 fm.
 \begin{figure}[tb]
  \begin{center}
\vspace{-1.2cm}
\includegraphics[width=11cm]{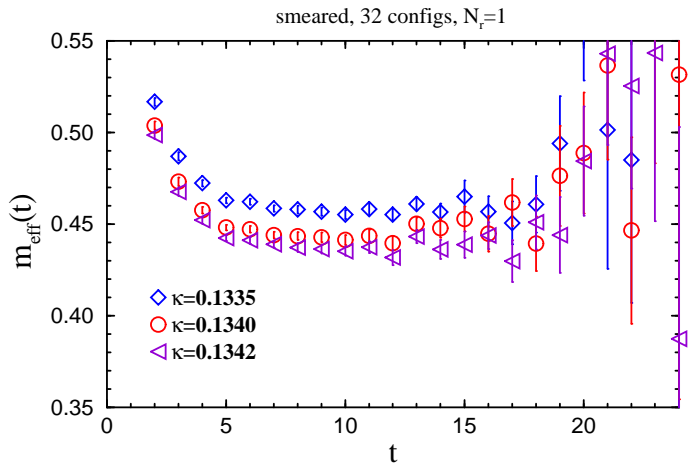}
\vspace{-1.7cm}
  \end{center}
   \caption{Plots of the effective mass of the 2-point function for
   $\kappa=0.1335$,
  01340, and 0.1342, with $(N_r,N_{t_0})=(1,48)$. We observe a clear
   plateau for $t\in [7,11]$} 
   \label{fig:2pt_meff}
\vspace{-0.4cm}
 \end{figure}
In this subsection we present our results for the 2-point correlators. 
Figure~\ref{fig:2pt_meff} presents effective mass plots of the 2-point
functions for $\kappa=0.1335$, 0.1340, and 0.1342. Fitting the 2-point
function with the fit range $t=\left[7,11\right]$, we obtain the
binding energy of the heavy-light meson with 0.2 \% accuracy as shown
in Fig.~\ref{fig:aE}. The square in Fig.~\ref{fig:aE} represents the
result of the Alpha collaboration~\cite{DellaMorte:2003mn} 
for the same value $\beta=6.0$ on a $16^3 \times 32$ lattice with 
the Schr\"odinger functional boundary conditions using 
on the order of a few hundred configurations. Although 
a simple comparison may not be meaningful, 
because of the different the kinematical setups used 
(for example, the volume, boundary conditions and operators), 
it is nevertheless noteworthy
that the all-to-all propagators yields much better precision with a much
smaller number of configurations, $N_{\rm conf}=32$. 
Extrapolating linearly in $(am_{\pi})^2$, we obtain
\begin{eqnarray}
a E = 0.408 \pm 0.002,
\end{eqnarray}
i.e., 0.5\% accuracy in the chiral limit.
%-----------------
\subsection{3-point correlators}	
%----------------
%
\begin{figure}[tbp]
\begin{center}
\vspace{-1.0cm}
 \includegraphics[width=10cm]{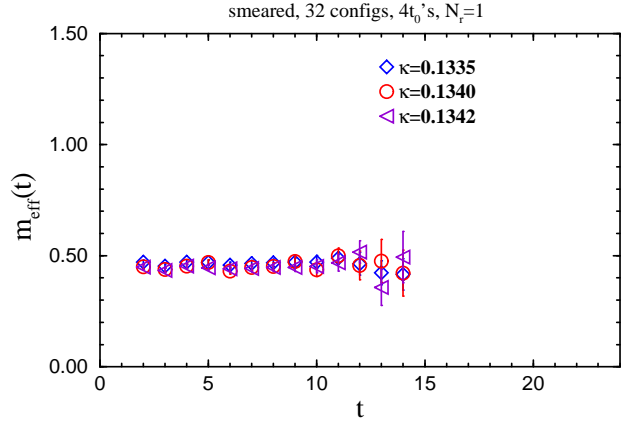}
\vspace{-1.5cm}
\end{center}
\caption{Plots of the effective mass of the 3-point correlation function for
$\kappa=0.1335$, 0.1340, and 0.1342 with 4 $t_0$ sources.}
   \label{fig:3pt_meff_4t0}
\end{figure}
\begin{figure}[tbp]
\begin{center}
\vspace{-0.8cm}
\includegraphics[width=10cm]{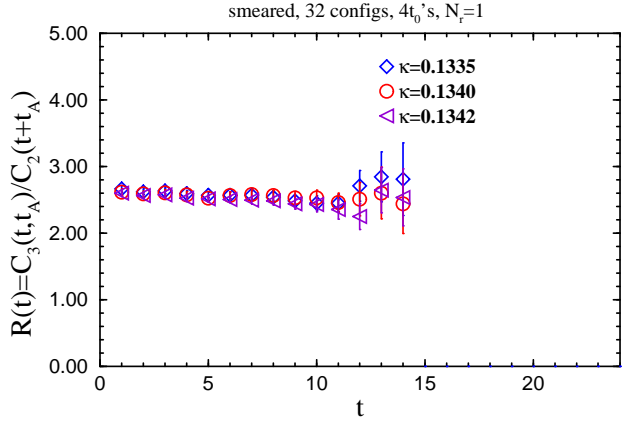}
\vspace{-1.5cm}
\end{center}
\caption{3-point/2-point correlator ratio  $R(t,t_A)$ for
$\kappa=0.1335$, 0.1340, and 0.1342  with 4 $t_0$ sources. We observe
   a clear plateau for $t\in [5,9]$.} 
   \label{fig:3pt_2pt_4t0}
\vspace{-0.2cm}
 \end{figure}
Figure~\ref{fig:3pt_meff_4t0} represents plots of the effective mass of the 3-point
correlators. We find that the plateau is reached for $t \geq 4$.
Figure~\ref{fig:3pt_2pt_4t0} displays the 3-point/2-point correlator
ratios. 
Applying constant fits of this ratio with fitting range $t=[5,9]$, we obtain
\begin{eqnarray}
R(t,t_A)=\frac{C_3(t,t_A)}{C_2(t+t_A)}= 2.56(5) \mbox{ for $\kappa=0.1335$},\\
R(t,t_A)=\frac{C_3(t,t_A)}{C_2(t+t_A)}= 2.55(5) \mbox{ for $\kappa=0.1340$},\\
R(t,t_A)=\frac{C_3(t,t_A)}{C_2(t+t_A)}= 2.50(4) \mbox{ for $\kappa=0.1342$},
\end{eqnarray}
where each result has 2\% accuracy.
The bare lattice operator for the light-light axial vector current
should be renormalized as
\begin{eqnarray}
A_i &=& 2\kappa Z_A (1+b_A m_q) ( A_i^{\rm lat} + c_A \partial_i P). 
\end{eqnarray}
The second term here does not contribute after summing over all spatial
positions. 
The quantity $Z_A$ has been determined nonperturbatively by the Alpha collaboration. 
For $\beta=6.0$, they obtained $Z_A=0.79$. The quantitly $b_A$ was also obtained by
combining the  
results of $b_V$ from the Alpha collaboration~\cite{Sint:1997jx}
and $b_A-b_V$ from Battacharya et al.
~\cite{Bhattacharya:2000pn} for $\beta=6.0$. Using the above
formula, the $B^*B\pi$ coupling can be obtained as 
\begin{eqnarray}
\hat{g}_{\infty} = 2\kappa Z_A (1+b_A m_q) R.
\end{eqnarray}
\begin{figure}[tbp]
\begin{center}
\vspace{-1.6cm}
 \includegraphics[width=11cm]{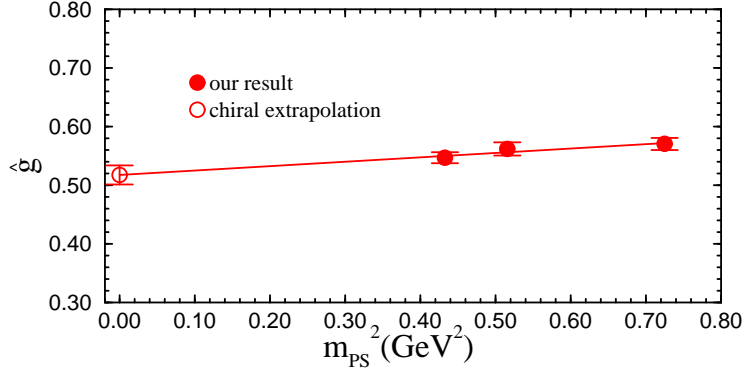}
\vspace{-1.8cm}
\end{center}
\caption{$\hat{g}_{\infty}$ for $\kappa=0.1335$, 0.1340, and  0.1342, 
together with the result in the chiral limit obtained by linear extrapolation
in $(am_{\pi})^2$.} 
\label{fig:chiral_g}
\vspace{-0.2cm}
\end{figure}
Figure~\ref{fig:chiral_g} displays our results the $\hat{g}_{\infty}$ as a
function of $(a m_{\pi})^2$. Taking the chiral limit by
extrapolating linearly in $(a m_{\pi})^2$, we obtain 
$\hat{g}_{\infty}$ in the chiral limit as
\begin{eqnarray}
\hat{g}_{\infty} = 0.517 \pm 0.016,  & &  \mbox{ \   (for $\beta=6.0$)}
\end{eqnarray}
with 3\% statistical error.
This result can be compared with the result on $16^3\times 32$
lattices for $\beta=6.0$ using the $O(a)$-improved Wilson fermion with
$c_{\rm SW}=1.769$  by Abada et al~\cite{Abada:2003un}. 
Their result in the chiral limit is $\hat{g}_{\infty} = 0.45(4)$, which
is consistent with ours  within 2$\sigma$.  We note that their
results for finite quark masses have 4--6\% statistical errors giving
9\% accuracy in the chiral limit with on the order of 100 configurations,
whereas ours have 2\% statistical error for finite quark masses giving
3\% accuracy in the chiral limit with 32 configurations. It is thus seen that are obtain
a significant improvement in statistical accuracy using the all-to-all
propagators.   
%%%%%%%%%%%%%%%%%%%%%%%%%%%%%%%%%%%%%%%%%%%%%%%%%%%%%
%
\section{Conclusions}
In this paper, we have studied the feasibility of the all-to-all propagator
method for precise computation of the $B^*B\pi$ coupling,
$\hat{g}_{\infty}$. Determining 100 eigenmodes, we find that the
low-modes dominate the correlators, and therefore the low-mode averaging in
the all-to-all propagator reduces the noise significantly.  

For reference we, list the numerical cost estimate 
using  1 CPU on our vector supercomputer in Table~\ref{tab:cost}.
\begin{table}[tb]
\caption{Computational cost per configuration for 2-point and 3-point correlators for $\kappa=0.1335$.}
\begin{center}
\begin{tabular}{lllll}
\hline
\hline
\multicolumn{5}{c}{2-point correlator}\\
$N_{\rm ev}$ & $N_{t_0}$ & low      &  high      & total \\
\hline
 0       &  4        & -        &  30 min.   & 30 min\\
 20      &  4        & 7 min.   &  35 min.   & 40 min\\
100      &  4        & 35 min.  &  35 min.   & 70 min\\
100      &  48       & 35 min.  &  7 h 25 min & 8 h\\
\hline
\hline
\multicolumn{5}{c}{3-point correlator}\\
$N_{\rm ev}$ & $N_{t_0}$ & low-low  &  low-high, high-low, high-high & total \\
\hline
100      &  4       & 40 min.  &  $\sim$ 32 h & $\sim$ 33 h\\
\hline
\end{tabular}
\end{center}
\label{tab:cost}
\vspace{-0.2cm}
\end{table}
For 2-point heavy-light meson correlators in quenched QCD, 8
hours are required per configuration using the all-to-all propagator with
$N_{\rm ev}=100$ and $(N_r,N_{t_0})=(1,48)$, while this time becomes 10 hours if the
time for the eigenmode calculation is included. If the equivalent computational cost
were used for the ordinary method with a single source timeslice, one would
have to compute correlators with 64--80 times more configurations,
which could reduce the error by a factor of 8--9. On the other
hand, the errors of the 2-point correlator for 32 configuration  
are a factor of 0.08--0.10 smaller than those obtained with the ordinary method. Therefore,
there is a marginal gain with the all-to-all propagator. We presume that
the situation would be similar for 3-point correlators, although we
have not made a comparison of the results obtained using all-to-all propagators
and those obtained with the ordinary method. 

However, the essential aim of this study is not to show that we have a
marginal  gain with the all-to-all propagator in quenched QCD. What is
important is that our result in quenched QCD suggests that there is
a significant gain in unquenched QCD, in which case
generating configurations is much more costly than
the measurements. Also, the fact that only 32
independent gauge configurations are sufficient for obtaining very
precise results with a few percent accuracy is encouraging.  
%
%----------------------------------------------------
\section*{Acknowledgements}
We would like to thank S.~Hashimoto and T.~T.~Takahashi for fruitful
discussions and comments. 
We would also like to thank T.~Goto, S.~Kawai and T.~Kimura  for useful
comments. The numerical calculations were carried out on the
vector supercomputer NEC SX-8 at YITP, Kyoto University, and a supercomputer NEC SX-5 at Research Center for Nuclear Physics, Osaka University. 
The authors also thank the members of YITP, Kyoto University, where this work was
initiated during  
the  YITP workshop ``Actions and symmetries in lattice gauge theory''
YITP-W-05-25. H.~M. and T.~O. are supported by Grants-in-Aid from the
Ministry of Education of Japan (No. 16740156 and Nos. 1315213, 16540243
respectively).  
%\appendix
%\section{First Appendix} %Empty argument \section{} yields `Appendix'. 
%
%\section{Second Appendix}
%\appendix
\setcounter{section}{1}
\renewcommand{\thesection}{\Alph{section}}
\renewcommand{\theequation}{\thesection$\cdot$\arabic{equation}}
\section*{Appendix}
In this appendix we give the definitions of the low-low, 
low-high, high-high, and high-low mode correlators,
$C_{ll}$, $C_{lh}$, $C_{hh}$, and $C_{hl}$, where  
the first (second) index of $C_{q_1q_2}$ specifies
the quark propagator from $t_A$ to $t_{B^*}$
(from $t_B$ to $t_A$).
The low-low contribution to the 3-pt functions is given 
by 
\begin{eqnarray}
C_{ll}(t_{B^*},t_A,t_B)
&=& \frac{1}{V}
        \sum_{i,j=1}^{N_{\rm ev}} \frac{1}{\lambda_i\lambda_j}
        \sum_{\vec{x},\vec{y},\vec{z}} \sum_{\vec{r},\vec{w}}
      {\rm Tr}\left\{
      \gamma_5 S_b^\dag(\vec{x},t_{B^*};t_B)
       \delta(\vec{y}-\vec{x}) \gamma_5 \Gamma_{B^*_\alpha}
   \right.
\nonumber\\
 & & \times\, \left[ v^{(i)}(\vec{x}+\vec{r},t_{B^*})\otimes
           v^{(i)\dag}(\vec{z},t_{A_i}) \right]
          \Gamma_{A_\alpha}
\nonumber\\
 & & \left. \times\,
      \left[ v^{(j)}(\vec{z},t_{A_i})\otimes
             v^{(j)\dag}(\vec{y}+\vec{w},t_B) \right]
       \tilde{\Gamma}_B
      \phi'(\vec{r}) \phi(\vec{w})
  \right\} 
\label{eq:3pt_ll} \\
 &=& \frac{1}{V} \sum_{i,v}^{N_{\rm ev}} \frac{1}{\lambda_i\lambda_j}
     \left\{  B^{ij}_\alpha(t_{B^*},t_B) A^{ij}_\alpha(t_A) \right\},
\label{eq:3pt_ll2}
\end{eqnarray}
where $\alpha=1,2,3$ and
\begin{eqnarray}
  A^{ij}_\alpha(t_A) &=&
    \sum_{\vec{z}} v^{(i)\dag}(\vec{z},t_A)\Gamma_{A_\alpha}
                   v^{(j)}(\vec{z},t_A) , \\
  B^{ij}_\alpha(t_{B^*},t_B) &=&
    \sum_{\vec{x}} v_\phi^{(j)\dag}(\vec{x},t_B)
        \Gamma'_B  S_b^{\dag}(\vec{x},t_{B^*};t_B)
        \Gamma'_{B_\alpha^*} v_{\phi'}^{(i)}(\vec{x},t_{B^*}) ,
\end{eqnarray}
with $\Gamma_{A\alpha}=\gamma_\alpha$, 
 $\Gamma'_B = \tilde{\Gamma}_B \gamma_5 = -\gamma_5$, 
and $\Gamma'_{B_\alpha^*} = \gamma_5 \Gamma_{B_\alpha^*}
 = \gamma_5\gamma_\alpha$.
%$A^{ij}_\alpha(t_A)$ and $B^{ij}_\alpha(t_{B^*},t_B)$ can be computed
%separately.

The low-high contribution to the 3-pt functions are given 
by 
\begin{eqnarray}
C_{lh}(t_{B^*},t_{A},t_B)
 &=& \frac{1}{V} 
        \sum_{\vec{x},\vec{y},\vec{z}} \sum_{\vec{r},\vec{w}}
    {\rm Tr}\Big\{ S_b(\vec{y},t_B;\vec{x},t_{B^*})
              \Gamma_{B^*_\alpha}
              \bar{Q}_0(\vec{x}+\vec{w},t_{B^*};\vec{z},t_A)
   \nonumber\\
 & & \times\,
              \Gamma_{A_\alpha}
              \bar{Q}_1(\vec{z},t_A;\vec{y}+\vec{w},t_B)
              \tilde{\Gamma}_B
              \phi(\vec{r}) \phi(\vec{w})
      \Big\}
\label{eq:3pt1_lh_1} \\
 &=& \frac{1}{V} \frac{1}{N_r} 
        \sum_{i=1}^{N_{\rm ev}} \frac{1}{\lambda_i}\sum_{r,j}
        \sum_{\vec{x},\vec{y},\vec{z}} \sum_{\vec{r},\vec{w}}
      {\rm Tr} \Big\{
      \gamma_5 S_b^\dag(\vec{x},t_{B^*};t_B)
       \delta(\vec{y}-\vec{x}) \gamma_5 \Gamma_{B^*_\alpha}
       \nonumber\\
 & &  \times
      \left[ v^{(i)}(\vec{x}+\vec{r},t_{B^*})\otimes
             v^{(i)\dag}(\vec{z},t_{A}) \right]
      \Gamma_{A_\alpha}
\nonumber\\
 & &  \times 
       \left[  \psi_{[r]}^{\prime(j)}(\vec{z},t_{A_i}) \otimes
              \eta_{[r]}^{(j)\dag}(\vec{y}+\vec{w},t_B) \right]
       \tilde{\Gamma}_B
      \phi(\vec{r}) \phi(\vec{w})
  \Big\}
\nonumber\\
 &=& \frac{1}{V} \frac{1}{N_r} 
        \sum_{i=1}^{N_{\rm ev}} \frac{1}{\lambda_i}\sum_{r,j}
       B^{ij}_{[r]\alpha}(t_{B^*},t_B) \cdot
       A^{ij}_{[r]\alpha}(t_A),
\label{eq:3pt1_lh_2}
\end{eqnarray}
where
\begin{eqnarray}
  A^{ij}_\alpha(t_A) &=&
    \sum_{\vec{z}} v^{(i)\dag}(\vec{z},t_A)\Gamma_{A_\alpha}
                   \psi^{(j)}_{[r]}(\vec{z},t_A) , \\
  B^{ij}_\alpha(t_{B^*},t_B) &=&
    \sum_{\vec{x}} \eta_{[r]\phi}^{(j)\dag}(\vec{x},t_B)
        \Gamma'_B  S_b^{\dag}(\vec{x},t_{B^*};t_B)
        \Gamma'_{B_\alpha^*} v_{\phi}^{(i)}(\vec{x},t_{B^*}).
\end{eqnarray}
\smallskip
While we repeatedly use the symbols $A^{ij}_\alpha(t_A)$
and $B^{ij}_\alpha(t_{B^*},t_B)$, we anticipate there will be
no confusion, since they are valid only in
the expression just before they are defined.

In the case of $C_{hl}$, we set the source of the noisy
estimator to $t_{B^*}$.
Since $Q_1$ is hermitian, we have
\begin{equation}
 \bar{Q}_1(x,z) = \bar{Q}_1(z,x)^\dag
  = \frac{1}{N_r}\sum_{r,i} \left[
        \psi^j_{[r]}(z) \otimes \eta^{j\dag}_{[r]}(x)
                   \right]^\dag
  = \frac{1}{N_r}\sum_{r,i} \left[
        \eta^j_{[r]}(x) \otimes \psi^{j\dag}_{[r]}(z)
                   \right] .
\end{equation}
This leads to
\begin{eqnarray}
C_{hl}(t_{B^*},t_{A},t_B)
 &=& \frac{1}{V} 
        \sum_{\vec{x},\vec{y},\vec{z}} \sum_{\vec{r},\vec{w}}
    {\rm Tr}\Big\{ S_b(\vec{y},t_B;\vec{x},t_{B^*})
              \Gamma_{B^*_\alpha}
              \bar{Q}_1(\vec{x}+\vec{w},t_{B^*};\vec{z},t_A)
   \nonumber\\
 & & \times\, \Gamma_{A_\alpha}
              \bar{Q}_0(\vec{z},t_A;\vec{y}+\vec{w},t_B)
              \tilde{\Gamma}_B
              \phi(\vec{r}) \phi(\vec{w})
      \Big\}
\label{eq:3pt1_hl_1} \\
 &=& \frac{1}{V}
     \frac{1}{N_r}\sum_{j=1}^{N_{\rm ev}} \frac{1}{\lambda_j}\sum_{r,i}
        \sum_{\vec{x},\vec{y},\vec{z}} \sum_{\vec{r},\vec{w}}
      {\rm Tr}\Big\{
      S_b(\vec{x},t_B;t_{B^*})
       \delta(\vec{y}-\vec{x})\Gamma_{B^*_\alpha}
      \nonumber\\
 & &  \times\,
      \left[  \eta^{(i)}_{[r]}(\vec{x}+\vec{r},t_{B^*})
         \otimes \psi^{(i)\dag}_{[r]}(\vec{z},t_A) \right]
      \Gamma_{A_\alpha}
 \nonumber\\
 & &  \times\,
     \left[  v^{(j)}(\vec{z},t_{A_i}) \otimes
              v^{(j)\dag}(\vec{y}+\vec{w},t_B) \right]
       \tilde{\Gamma}_B
      \phi(\vec{r}) \phi(\vec{w})
  \Big\}, \hspace{0.5cm}
\nonumber\\
 &=& \frac{1}{V} \frac{1}{N_r} 
        \sum_{i=1}^{N_{\rm ev}} \frac{1}{\lambda_i}\sum_{r,j}
       B^{ij}_{[r]\alpha}(t_{B^*},t_B) \cdot
       A^{ij}_{[r]\alpha}(t_A),
\label{eq:3pt1_hl_2}
\end{eqnarray}
where
\begin{eqnarray}
  A^{ij}_\alpha(t_A) &=&
    \sum_{\vec{z}} \psi^{(i)\dag}_{[r]}(\vec{z},t_A)
                  \Gamma_{A_\alpha}
                   v^{(j)}(\vec{z},t_A) , \\
  B^{ij}_\alpha(t_{B^*},t_B) &=&
    \sum_{\vec{x}} v_{\phi}^{(j)\dag}(\vec{x},t_B)
        \tilde{\Gamma}_B  S_b(\vec{x},t_B;t_{B^*})
        \Gamma_{B_\alpha^*} \eta_{[r]\phi}^{(i)}(\vec{x},t_{B^*}).
\end{eqnarray}
Recall that 
$\Gamma_{B^*_\alpha}=\gamma_\alpha$,
$\Gamma_{A_\alpha}=\gamma_\alpha$, and $\tilde{\Gamma}_B=-1$
($\alpha=1,2,3$).
\smallskip

The high-high part, $C_{hh}^{(1)}$, is evaluated using
the noisy estimator with a source vector put on the time
slice $t_B$:
\begin{eqnarray}
C_{hh}(t_{B^*},t_{A},t_B)
 &=& \frac{1}{V} 
        \sum_{\vec{x},\vec{y},\vec{z}} \sum_{\vec{r},\vec{w}}
    {\rm Tr}\Big\{ S_b(\vec{y},t_B;\vec{x},t_{B^*})
              \Gamma_{B^*_\alpha}
              \bar{Q}_1(\vec{x}+\vec{w},t_{B^*};\vec{z},t_A)
   \nonumber\\
 & & \times\, \Gamma_{A_\alpha}
              \bar{Q}_1(\vec{z},t_A;\vec{y}+\vec{w},t_B)
              \tilde{\Gamma}_B
              \phi(\vec{r}) \phi(\vec{w})
      \Big\}
\label{eq:3pt1_hh_1} \\
 &=& \frac{1}{V}
     \frac{1}{N_r} \sum_{r,i}
        \sum_{\vec{x},\vec{y},\vec{z}} \sum_{\vec{r},\vec{w}}
      {\rm Tr}\Big\{
      S_b(\vec{y},t_B;\vec{x},t_{B^*})
       \Gamma_{B^*_\alpha}
       \bar{Q}_1(\vec{x}+\vec{w},t_{B^*};\vec{z},t_A)
      \nonumber\\
 & &  \times\,
      \Gamma_{A_\alpha}
      \left[ \psi^{(i)}_{[r]}(\vec{z},t_A)
      \otimes \eta^{(i)\dag}_{[r]}(\vec{y}+\vec{w},t_B) \right]
      \tilde{\Gamma}_{B}
              \phi(\vec{r}) \phi(\vec{w}) .
\end{eqnarray}
Using the source method,
\begin{eqnarray}
  \Psi_{[r]\alpha}^{(i)}(\vec{x},t_{B^*};t_A)
  &=& \sum_{\vec{z}}
      \bar{Q}_1(\vec{x},t_{B^*};\vec{z},t_A)
      \Gamma_{A_\alpha}
      \psi^{(i)}_{[r]}(\vec{z},t_A)
\nonumber \\
  &=& \sum_{\vec{z}}
      D^{-1}(\vec{x},t_{B^*};\vec{z},t_A)
      \gamma_5 P_1 \Gamma_{A_\alpha}
      \psi^{(i)}_{[r]}(\vec{z},t_A)
\end{eqnarray}
can be obtained by solving the linear equation
$D\Psi=b$, with the source vector
$b=\gamma_5 P_1 \Gamma_{A_\alpha}
\psi^{(i)}_{[r]}(\vec{z},t_A)$.
Then $C_{hh}$ is obtained as
\begin{equation}
C_{hh}(t_{B^*},t_{A},t_B)
 = \frac{1}{V} \frac{1}{N_r} \sum_{r,i}
     \sum_{\vec{x}}
     \eta^{(i)\dag}_{[r]\phi}(\vec{x},t_B) 
     \tilde{\Gamma}_{B}
     S_b(\vec{x},t_B;t_{B^*})
     \Gamma_{B^*_\alpha}
     \Psi_{[r]\alpha,\phi}^{(i)}(\vec{x},t_{B^*};t_A) .
\end{equation}
We must solve $3\times N_{t_A}$ linear equations
per noise vector,
where $N_{t_A}$ is the number of sampled time slices
$t_A$.

\end{document}